\PassOptionsToPackage{dvipsnames}{xcolor}

\documentclass[conference]{IEEEtran}
\IEEEoverridecommandlockouts
\usepackage{cite}
\usepackage{amsmath,amssymb,amsfonts,amsthm}
\usepackage[endLComment=]{algpseudocodex}
\usepackage[ruled]{algorithm}
\usepackage{etoolbox}
\patchcmd{\subsubsection}{\itshape}{\bfseries}{}{}
\usepackage{graphicx}
\usepackage{subcaption}
\usepackage[normalem]{ulem}
\usepackage{textcomp}
\usepackage[dvipsnames]{xcolor}
\usepackage{comment}
\usepackage{soul}
\usepackage{multirow}
\usepackage[inline]{enumitem}
\usepackage{hyperref}
\usepackage[capitalise]{cleveref}
\usepackage{siunitx}
\def\BibTeX{{\rm B\kern-.05em{\sc i\kern-.025em b}\kern-.08em
    T\kern-.1667em\lower.7ex\hbox{E}\kern-.125emX}}

\theoremstyle{definition}

\theoremstyle{theorem}

\theoremstyle{corollary}

\theoremstyle{lemma}

\usepackage{xargs}
\usepackage{sidecap}

\usepackage[colorinlistoftodos,prependcaption,textsize=tiny]{todonotes}

\algnewcommand\algorithmicforeach{\textbf{for each}}
\algdef{S}[FOR]{ForEach}[1]{\algorithmicforeach\ #1\ \algorithmicdo}

\setlist[enumerate]{nosep}

\begin{document}


\title{MassiveGNN: Efficient Training via Prefetching \\ for Massively Connected Distributed Graphs}

\author{
    \IEEEauthorblockN{Aishwarya Sarkar\IEEEauthorrefmark{1}, Sayan Ghosh\IEEEauthorrefmark{2}, Nathan R. Tallent\IEEEauthorrefmark{2}, Ali Jannesari\IEEEauthorrefmark{1}}
    \IEEEauthorblockA{\IEEEauthorrefmark{1}Iowa State University, Ames, IA, USA, Email: \{asarkar1, jannesari\}@iastate.edu}
    \IEEEauthorblockA{\IEEEauthorrefmark{2}Pacific Northwest National Laboratory, Richland, WA, USA, Email: \{sayan.ghosh, tallent\}@pnnl.gov}
}
\maketitle

\begin{abstract}
Graph Neural Networks (GNN) are indispensable in learning from graph-structured data, yet their rising computational costs, especially on massively connected graphs, pose significant challenges in terms of execution performance. To tackle this, distributed-memory solutions such as partitioning the graph to concurrently train multiple replicas of GNNs are in practice. However, approaches requiring a partitioned graph usually suffer from communication overhead and load imbalance, even under optimal partitioning and communication strategies due to irregularities in the neighborhood minibatch sampling. 

This paper proposes practical trade-offs for improving the sampling and communication overheads for representation learning on distributed graphs (using popular GraphSAGE architecture) by developing a parameterized continuous \emph{prefetch} and \emph{eviction} scheme on top of the state-of-the-art Amazon DistDGL distributed GNN framework, demonstrating about 15--40\% improvement in end-to-end training performance on the National Energy Research Scientific Computing Center's (NERSC) Perlmutter supercomputer for various OGB datasets. 
\end{abstract}

\thispagestyle{plain}
\pagestyle{plain}

\section{Introduction}\label{sec:intro}
Distributed Graph Neural Networks (GNNs) have emerged as a leading paradigm for machine learning on massive graphs, demonstrating success in domains such as molecular biology, social networks, recommendation systems, traffic forecasting, and climate models~\cite{zhou2020graph, wu2020comprehensive, jiang2022graph, zhang2021graph}. 
Following the norms of established deep learning (DL)-based machine learning (ML) domains such as Natural Language Processing (NLP) and Computer Vision (CV), minibatch-driven training has emerged as a practical, memory-efficient solution with strong convergence guarantees~\cite{bottou2010large}. However, the rising diversity and complexity of massive real-world graph datasets necessitate interventions for conducting GNN training efficiently at scale~\cite{shao2024distributed}, while adopting the current best-practices in minibatch training. Minibatch-driven training for graph-structured data poses distinct challenges compared to other ML domains, such as CV, NLP, etc., due to the inherent connections within the graphs, implying interdependencies that must be captured.    

GNNs are primarily designed to tackle node prediction, link prediction, and graph prediction. Within the scope of this work, we specifically focus on the \emph{node classification} task that entails using a GNN to classify nodes within a graph using their interconnections (edges) and the node features. Distributed-memory training on graphs typically involves partitioning a large graph using advanced Graph Partitioning (GP) algorithms~\cite{bulucc2016recent,karypis1998fast} to distribute the induced subgraphs among workers\slash Processing Elements (PE). In GNNs, ``messages'' generated on a ``neighborhood'' of edges are aggregated on vertices\slash nodes they are incident on (this is referred as ``message passing''). An effect of the ``message passing'' is the ``neighborhood explosion'' phenomenon~\cite{zeng2019graphsaint} in which neighborhood inspection of an arbitrary node spans the entire graph, thereby raising computational expenses significantly due to the multiplicative effect of enhanced connections. Layer \emph{sampling} \cite{balın2023layerneighbor} was proposed to mitigate this issue, which considers a random sample of the neighborhood subgraph of nodes; sampling is recursively invoked for as many layers there are in the GNN model. In distributed-memory GNNs, a sampled neighborhood can be distributed across several partitions, requiring repetitive communication to collect the features before training can resume for every minibatch iteration, undermining the efficiency gains from distributed-memory training. 

Prior work to alleviate communication issues of GNN training can be broadly classified into \emph{communication optimization} or \emph{communication avoidance} strategies. Communication optimization strategies try to reduce the volume of communication, e.g., via data compression methods~\cite{9835576}, reducing communication rounds, prefetching minibatches in advance~\cite{lin2020pagraph}, etc. In contrast, communication avoidance approaches ignore neighboring ``halo'' nodes from consideration, at the expense of model performance~\cite{ramezani2022learn}, and are yet to be widely accepted. A straightforward \emph{communication optimization} might involve retaining (``caching'') currently sampled nodes, but it may not yield benefits in the long run (over minibatches), since sampling is non-deterministic (sampling is random walk based, stochastic in nature). Another promising \emph{communication optimization} is prefetching successive minibatches (with duplicate node resolution), but it requires careful memory management and data distribution strategies. \cite{tripathy2024distributed, hoang2023batchgnn} 

\begin{figure}[!ht]
    \centering
    \includegraphics[width=0.85\columnwidth]{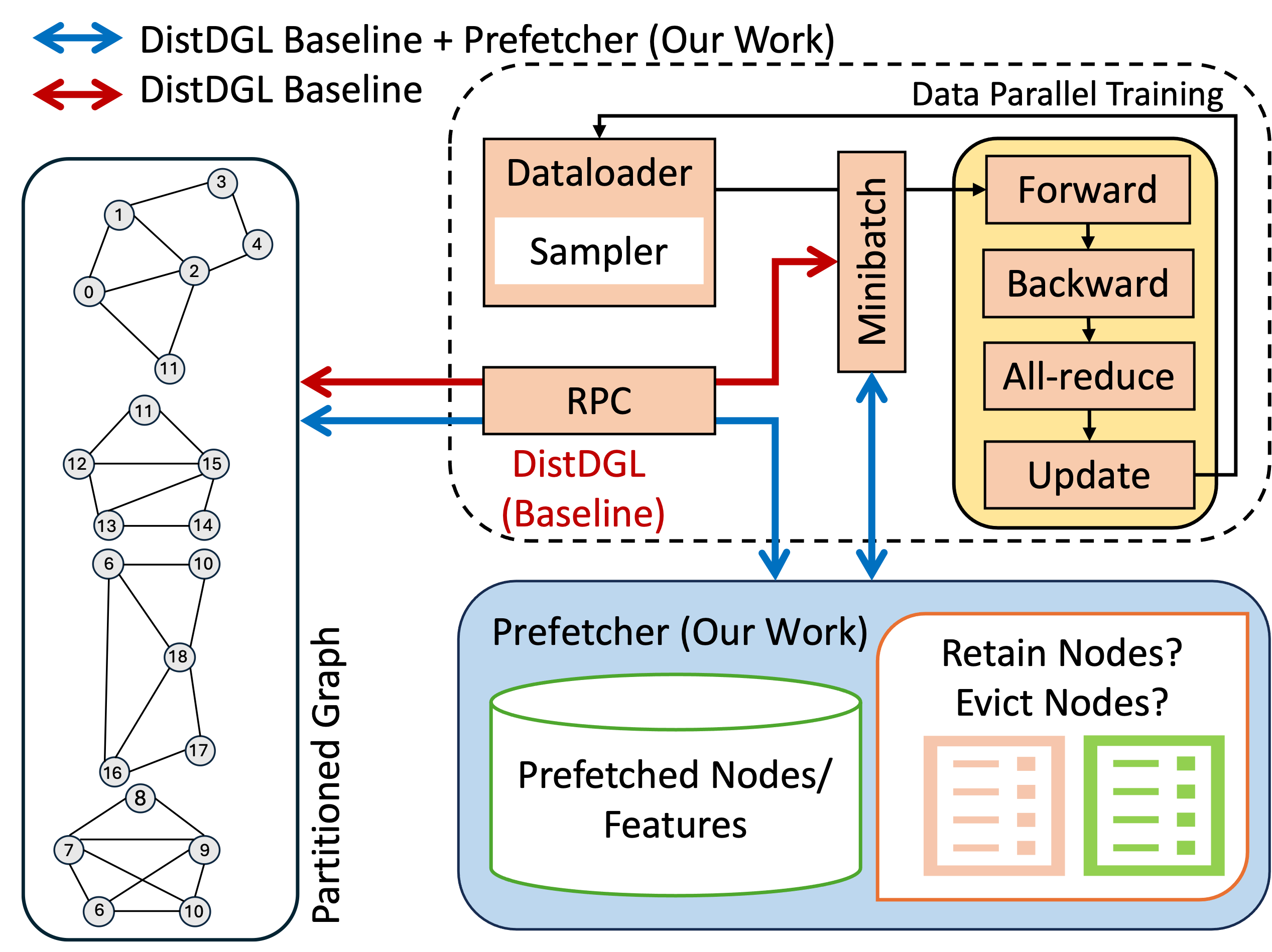}
    \caption{Extending DistDGL with our continuous prefetch and eviction scheme.}
    \label{fig:contributions}
\end{figure}

Our solution to mitigate the rising communication overheads for learning on massively connected graphs involves a \emph{continuous} prefetch and eviction process to tally and move nodes (features) into and out of a preallocated buffer for preparing the next minibatch, overlapped with the learning process invoked on the current minibatch, implemented within the state-of-the-art Amazon DistDGL platform for GNN training~\cite{zheng2020distdgl} as shown in Fig.~\ref{fig:contributions}. Since CPU resources are relatively underutilized in contemporary GPU-focused training (with \#PEs involved in training equivalent to the \#GPUs, leaving most of the CPU cores idle), we exploit multicore CPUs in asynchronously overlapping our prefetching mechanism with data-parallel training. Ideally, we can achieve a perfect overlap (implying zero overhead), leading to about $15\%-40\%$ (with up to $85\%$ in OGB-arxiv) reduction in the RPC communication compared to DistDGL (by actively utilizing the prefetched nodes across minibatches), which improves the overall training times by the same margin, as observed on NERSC Perlmutter. 

Our contributions are:
\begin{enumerate}[itemsep=0pt]
    \item Parameterized continuous prefetch and eviction strategies considering pre-initialized ``prefetch'' buffers to retain frequently accessed node features.
    \item Implement prefetch and eviction scheme within the state-of-the-art DistDGL GNN training implementation.
    \item Formally analyze the impact of overlap for the next minibatch preparation and training processes.
    \item Detailed performance assessments (baseline CPU\slash GPU performance, prefetching utilization, communication reduction, trade-offs, etc.) of training the GraphSAGE model on NERSC Perlmutter using diverse OGB datasets.
\end{enumerate}
This paper is organized as follows: \S\ref{sec:back} provides a brief background on DistDGL. We discuss the prefetching and eviction methodology in \S\ref{sec:method}. In \S\ref{sec:eval}, we conduct detailed performance analysis. \S\ref{sec:summary} recapitulates relevant contributions.

\section{DistDGL framework}\label{sec:back}
DistDGL (Distributed Deep Graph Library) \cite{zheng2020distdgl} is a framework for training GNNs on distributed-memory systems, which internally uses PyTorch's Distributed Data Parallel (DDP) training framework for dense tensor operations (e.g., aggregation of GNN parameters for synchronous stochastic gradient descent in training). 
\begin{figure}[!ht]
    \centering
    \includegraphics[width=0.8\linewidth]{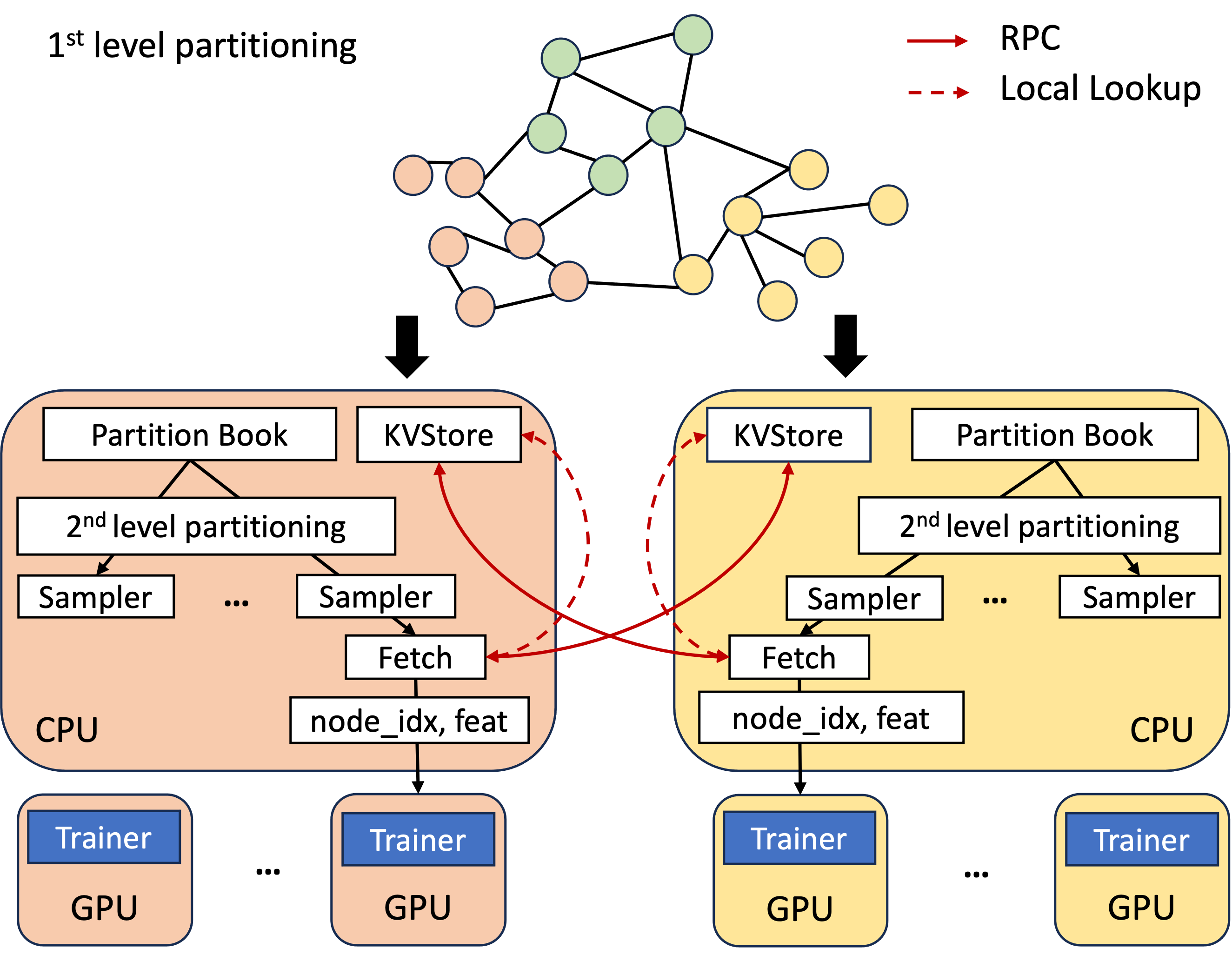}
    \caption{DistDGL architecture showcasing multiple samplers and trainers across distinct partitions, each represented by a different color.}
    \label{fig:distdgl}
\end{figure} 
The overall architecture of DistDGL is based on the client-server model, where each node in the cluster acts as a server. DistDGL spawns client PEs as \emph{trainers} and supports CPU\slash GPU platforms.

\paragraph*{Partitioning} DistDGL implements two levels of partitioning as shown in Fig.~\ref{fig:distdgl}. The \emph{first level} of partitioning occurs offline when the entire graph is divided into induced subgraphs (vertex-based, considering nodes with several ``cut'' edges or ``halo'' nodes, spanning adjacent partitions) and stored on the filesystem. In the \emph{second level} of partitioning, DistDGL re-distributes the nodes of a partition among the trainer processes. 

\paragraph*{Communication} 
Each trainer PE instantiates distributed \texttt{DataLoader} utility, which performs \emph{sampling} from the local partition (considering halo nodes) and returns a set of node indices (indices can determine whether a node is available in local or a remote partition), used to request associated node features. Local node features are retrieved from the key-value store (KVStore), whereas halo node features are requested from other servers via RPC (Remote Procedure Call). 
 
\paragraph*{Sample prefetching} 
DistDGL supports lazy feature prefetching, which supports pre-loading user-specified node features concurrently with model computation, potentially hiding some CPU-GPU communication (still requires inter-node communication for each sampled remote node). DistDGL also allows for multiple samplers to ``pre-sample" future minibatches. However, this option is currently not functional in our baseline version of DistDGL\footnote{\url{https://github.com/dmlc/dgl/issues/5731}.}. While these options can partially achieve overlap with training, unlike this work, they do not optimize network communication by maintaining prefetched features across the minibatches.
\section{Related Work}\label{sec:related}
\noindent\textbf{Shared-memory Prefetching:} 
Recent works have explored prefetching in the context of single-node (shared-memory) systems; for instance, Pagraph \cite{lin2020pagraph} proposes a static degree-based GPU-caching for single-node GNN training. They also point out that a dynamic caching policy is not suitable for an on-GPU cache since all computations performed on GPU must be assembled into the GPU kernels and launched by the CPU. GNNLab \cite{10.1145/3492321.3519557} employs a GPU-based static feature caching policy for single-node, multi-GPU training where a pre-sampling-based caching policy is shown to outperform degree policy methods, introducing a hotness metric to select node features for caching. However, this approach relies on the assumption that the sampler will follow a similar pattern during the pre-sampling phase and the actual training, which may not be true due to the inherent non-determinism of the sampling process. Existing works such as MariusGNN \cite{waleffe2023mariusgnn} and \cite{ducati} have bypassed communication entirely by leveraging single-node memory hierarchy, focusing on CPU-GPU data movement bottleneck during minibatch generation.

\noindent\textbf{Distributed-memory Prefetching:} BGL \cite{liu2023bgl}, a distributed training system addresses the cross-partition communication bottleneck by proposing a dynamic two-tiered (CPU and GPU) caching mechanism. To achieve this despite the non-determinism in sampling algorithms, they introduced a BFS-based partitioning module with a proximity-aware ordering for sampling nodes, not supporting some widely used sampling algorithms. Existing works\cite{tripathy2024distributed, jiao2023pglbox} have proposed in-memory and out-of-core bulk-sampling strategies for communication avoidance, requiring special data structures. Our work is similar to BatchGNN~\cite{hoang2023batchgnn} in some respects. BatchGNN consolidates successive feature fetches into a single macrobatch operation. A macrobatch combines multiple minibatches, reducing communication by only needing to communicate once for the common remote nodes sampled in the current macrobatch. This approach enforces using a single trainer PE\slash node and dedicates all available cores to perform macrobatch preparation. Unlike this work, they do not maintain a prefetch buffer to track nodes across the macrobatches. BatchGNN also optionally caches pre-computed aggregation of neighbors for each node (with limitations) to reduce the computation/communication during training, significantly increasing their memory overhead. We are unable to compare performance, as BatchGNN is not publicly available, and results presented in the paper are not directly comparable to our 4-trainers\slash node setup.
\noindent\textbf{Prefetching Strategies:} In terms of effective prefetching strategies, AliGraph \cite{zhu2019aligraph} caches the outgoing neighbors of the \emph{important} vertices, based on a metric that considers the number of k-hop in and out neighbors per node. In contrast, Pagraph \cite{lin2020pagraph} employs a static caching scheme where nodes are pre-sorted offline by out-degrees, and highest-degree nodes are cached on GPUs, considering the currently available memory. Salient++ \cite{Kaler2023.SALIENT++} is built on PyTorch Geometric (PyG)~\cite{fey2019fast} and introduces a static caching policy with customized priority-based ordering of vertices within a partition, additionally enabling several samplers to optimize training. 
\section{Prefetching and Eviction}\label{sec:method}
\begin{figure}[h]
    \centering
    \includegraphics[width=\linewidth]{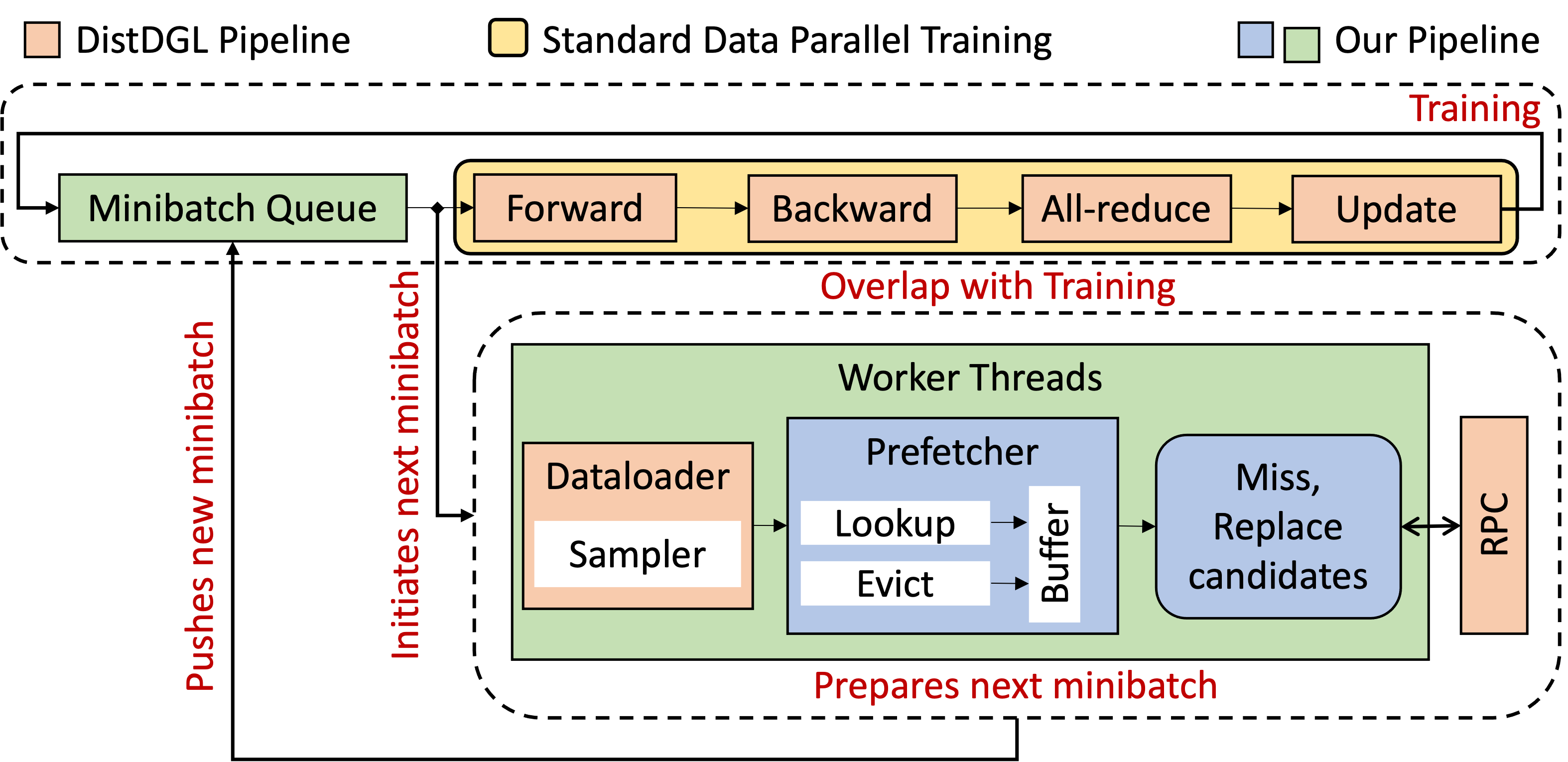}
    \caption{Workflow of proposed prefetch and eviction scheme.}
    \label{fig:workflow}
\end{figure}

We introduce our workflow in Fig.~\ref{fig:workflow} which highlights the key components of the prefetching and eviction pipeline, integrated within the DistDGL, which uses PyTorch's DDP training framework. We first discuss the relevant prefetching and eviction algorithms in the context of distributed training in \S\ref{ssec:method-algorithms}. Next, we discuss the scoring strategies underpinning our scheme in \S\ref{ssec:method-eviction}. In \S\ref{ssec:method-model}, we develop performance models to examine the impact of specific components. We introduce the \emph{hit rate} metric in \S\ref{ssec:method-metrics}, which is used to assess the quality of prefetching. Finally, we discuss trade-off scenarios for prefetching in \S\ref{ssec:method-tradeoff}. 

\subsection{Distributed Training Algorithms}\label{ssec:method-algorithms}
In this section, we discuss the distributed training algorithms, with our prefetching adaptations which are explained in the forthcoming sections. We begin by referring to common notations used in the algorithms as listed in Table~\ref{tab:notation}. 
\begin{table}[!ht]
\centering
\caption{Glossary of Notations}
\label{tab:notation}
\begin{tabular}{ll}
\hline
\textbf{Symbol} & \textbf{Description} \\ 
\hline
\emph{General} \\
\hline
$p$ & Partition number, \#partitions equals \#compute nodes \\
$i$ & Trainer processing element, (e.g., $i \in \#GPUs$) \\ 
$G(V, E)$ & Input graph with set of vertices ($V$) and edges ($E$) \\ 
$\Psi$ & GNN model (e.g., GraphSAGE) \\
$B$ & Per-trainer minibatch of features, labels, and blocks \\
\hline 
\emph{Per-partition} \\
\hline
$G_p(V_p,E_p)$ & Spatial graph and features in partition $p$  \\
$V_{p}^{h}$ & Remotely owned (halo) nodes in $V_{p}$ \\
$V_{p}^{l}$ & Locally owned nodes in $V_{p}$ ($V_{p} \supseteq  V_{p}^{h} \cup V_{p}^{l}$) \\
\hline 
\emph{Prefetch\slash Eviction} & \emph{(per-partition)}\\
\hline
$Q$ & Queue to store future minibatches\\
$\Delta, \alpha$, $\gamma$  & Eviction interval, threshold, rate of decay \\
$f^{h}_{p}$ & \% halo nodes to prefetch while initializing $\texttt{BUF}^{i}_{p}$ \\
$\texttt{BUF}^{i}_{p}$ & Prefetched halo nodes and features from $p$ by $i$\\
$V_{p}^{h|e}, V_{p}^{h|r}$ & Eviction and replacement candidate (halo) nodes \\
$S_E, S_A$ & Eviction\slash Access scores (updated per minibatch)\\
\hline 
\end{tabular}
\end{table}
\begin{algorithm}[!ht]
{\small
\caption{Our high-level distributed GNN training methodology with prefetching\label{algo:gnn-training}. Trainer 
\texttt{Dataloader} \textit{iterates} over specific training data (i.e., $iter(i, G_{p})$). 
Table~\ref{tab:notation} lists notations.}

\begin{algorithmic}[1]
\State Initialize model $\Psi$ parameters 
\State \Call{initialize\_prefetcher}{$i, G_p, f^{h}_{p}, S_{A}, S_{E}, \gamma, \Delta, \alpha$}
\For{$e \in range(\#epochs)$}
    \For{$step \in range(\#minibatches)$}
        \State $B \leftarrow Q.pop()$
        \If{$B$ is empty}
            \State $B \gets $\Call{prefetch\_with\_eviction}{$iter(i, G_{p}), step$}
        \EndIf
        \LComment{Overlap prefetching with training}
        \State \texttt{async} \Call{prepare\_next\_minibatch}{$Q,G_{p},step,i$}
        \LComment{Distributed Data Parallel Training}
        \Function{train\_ddp}{$\Psi, B, i$} 
            \State Perform forward pass
            \State Compute loss and perform backward pass
            \State Update parameters
        \EndFunction
        \State \texttt{Synchronize}
    \EndFor
\EndFor
\Procedure{initialize\_prefetcher}{$i, G_p, f^{h}_{p}, S_{A}, S_{E}, \gamma, \Delta, \alpha$}
    \State Setup $\texttt{BUF}_{p}^{i}$ and $S_E$ of size $\mathcal{O}(|V_{p}^{h}|.f_{p}^{h})$
    \State Initialize $\texttt{BUF}_{p}^{i}$ with top $f_{p}^{h}$\% of halo nodes by degree \Comment{RPC}
    \State Initialize $S_E[n]=1, S_A[n]=-1$ $\forall n \in \texttt{BUF}_{p}^{i}$
    \State Setup $S_A$ of size $\mathcal{O}(|V|)$
    \State Initialize $S_A[m]=0 \forall m \in V_{p}^{h} \setminus \texttt{BUF}_{p}^{i}$
    \State Initialize $\gamma$, $\Delta$, $\alpha$    
\EndProcedure
\Procedure{prepare\_next\_minibatch}{$Q, G_{p},step,i$}
    \State $B \gets$ \Call{prefetch\_with\_eviction}{$iter(i, G_{p}),step$}
    \State $Q.push(B)$ \Comment{Push next minibatch}
\EndProcedure
\end{algorithmic}
}
\end{algorithm}
\normalsize
Algorithm~\ref{algo:gnn-training} discusses the high-level GNN training algorithm using partitioned graph on several trainer processing elements. Training is conducted over a number of predetermined epochs and minibatch steps\slash iterations, with the trainer \texttt{DataLoader} fetching minibatches from the partitioned graph. Lines \#11--14 are typical DDP training across trainer PEs on compute nodes, which internally use collective operations to synchronize gradients (during backward pass) and update its local model's parameters. Before the training can begin, it is necessary to assemble the sampled node features from several partitions for the current minibatch. This is where we diverge from the baseline implementation (i.e., DistDGL), considering a pre-initialized dynamic prefetch buffer (associated parameters are initialized in the \Call{initialize\_prefetcher}{} function) to prepare the next minibatches asynchronously while the current batch is in training. \Call{initialize\_prefetcher}{} initializes $\texttt{BUF}^{i}_{p}$ by selecting a subset ($f_{p}^{h}\%$) of the high-degree remote nodes, predicated on the understanding that nodes with relatively higher number of connections, or degrees, are more likely to appear in random samples used during minibatch training. 
The \Call{prepare\_next\_minibatch}{} function returns the next minibatch, which is inserted into a queue for subsequent use. Unlike the baseline version, which has to wait until the sampled node (features) have been fetched (this process can invoke RPC for bulk communication) and copied, our implementation can begin training on subsequent minibatch steps immediately since the data already exists.   
\begin{algorithm}[!ht]
{\small
\caption{\textsc{prefetch\_with\_eviction}: High-level description of proposed \textit{prefetch} and \textit{eviction} steps to prepare the next minibatch, 
overlapped with data-parallel training on current minibatch. \label{algo:prefetch-evict}\\
\textbf{Input:} $G_{p}[step], step$ \\
\textbf{Output:} next minibatch, $B$}
\begin{algorithmic}[1]
\State $V_{p}^{h|l} \gets$ \textsc{NeighborSampler}($G_{p}[step]$) \Comment{Sampled neighbors}
\State $V_{p}^{l|s} \gets V_{p}^{h|l} \cap V_{p}^{l}$ \Comment{Local nodes in sampled minibatch}
\State $V_{p}^{h|s} \gets V_{p}^{h|l} \cap V_{p}^{h}$ \Comment{Halo nodes for prefetch and eviction}
\State $Hits \gets V_{p}^{h|s} \cap \texttt{BUF}_{p}^{i}$ \Comment{Halo nodes present in buffer}
\State $Misses \gets V_{p}^{h|s} \setminus \texttt{BUF}_{p}^{i}$ \Comment{Halo nodes absent in buffer}
\LComment{Start threads for parallel lookup (\textsc{NUMBA})}
\For{each $n \in \texttt{BUF}_{p}^{i}$ in parallel}
    \If{$n \notin V_{p}^{h|s}$}
    \State $S_E[n] = S_E[n] \cdot \gamma$ \Comment{Apply decay on unused}
    \EndIf
\EndFor
\State $B_{p}^{l} \gets$ $\forall n\in V_{p}^{l|s}$ \Comment{Copy local features}
\State $B_{p}^{h} \gets \texttt{BUF}_{p}^{i}[Hits]$ \Comment{Copy remote features from buffer}
\LComment{Lookup and evict period}
\If{$step=\Delta$} 
    \State $V_{p}^{h|e}, V_{p}^{h|r} \gets$ \Call{evict\_and\_replace}{$S_{E}, S_{A}, \alpha$}
    \State $F \gets$ Fetch features of $Misses$ and $V_{p}^{h|r}$ nodes \Comment{RPC}
    \State $\texttt{BUF}_{p}^{i}[V_{p}^{h|e}] = V_{p}^{h|r}$ \Comment{Evict \& replace}
    \State Update $\texttt{BUF}_{p}^{i}$ with $F[V_{p}^{h|r}]$ \Comment{Replace features}
    \State Update $B_{p}^{h}$ with $F[Misses]$ \Comment{Update minibatch}
    \State \Return $B_{p}^{h} \cup B_{p}^{l}$ \Comment{Return next minibatch}
\EndIf
\LComment{Start a thread to overlap fetching with access score update}
\State \texttt{async} \{$S_A[n]=S_A[n]+1$ $\forall n \in Misses$\} 
\State $B_{p}^{h}\gets$ Fetch features of $Misses$ \Comment{RPC}
\State \texttt{Synchronize}
\State \Return $B_{p}^{h} \cup B_{p}^{l}$ \Comment{Return next minibatch}
\Procedure{evict\_and\_replace}{$S_{E}$, $S_{A}$, $\alpha$}
\For{$n \in \texttt{BUF}_{p}^{i}$}
\LComment{Eviction and replacement assessment}
    \If{$S_{E}[n] < \alpha$} 
    \State $V_{p}^{h|e} = V_{p}^{h|e} \cup n$ \Comment{Eviction nodes}
    \If{$m = \mathit{max}(S_{A})$ is not empty}
    \State $V_{p}^{h|r} = V_{p}^{h|r} \cup m$ \Comment{Replacement nodes}
    \LComment{Reset to be prefetched node $m$}
    \State $S_A[m] = -1$ 
    \EndIf
    \EndIf
\EndFor
\State \Return $V_{p}^{h|e}, V_{p}^{h|r}$ \Comment{Return Eviction\slash Replacement nodes}
\EndProcedure
\end{algorithmic}
}
\end{algorithm}
\normalsize
Since GNN sampling is non-deterministic, it is impossible to predict the neighboring nodes returned by the sampler in every minibatch. Hence, we try to retain the node features for predefined periods in our prefetch buffer (maintaining one buffer per trainer PE) across a range of minibatches, recording past sampling trends by maintaining a scoreboard to track suitable nodes for retention. This is shown in Algorithm~\ref{algo:prefetch-evict} (\Call{prefetch\_with\_eviction}{}), which is a vital part of our next minibatch preparation strategy. 

Overall, we identify the ``halo'' or remotely owned nodes returned by the sampler, and try to retain them locally, such that next time they are needed\slash sampled, RPC communication can be avoided. We maintain two scores (invoking parallelism and asynchrony whenever possible), one to determine which nodes to evict from the prefetch buffer and another to figure out the nodes the sampler returned on successive minibatches but were not found on the prefetch buffer (this constitutes a ``miss'', whereas if a halo node is found in the prefetch buffer, it is a ``hit''). This is represented in lines \#1--11 of Algorithm~\ref{algo:prefetch-evict}. Since the act of ``evicting'' requires extra work (lookup and score update), it is only incorporated in predefined periods, which is represented in lines \#12--19. The  \Call{evict\_and\_replace}{} function (lines \#25--34) performs this assessment by marking the current nodes in the prefetch buffer whose scores are below a threshold ($\alpha$). 

\subsection{Scoring strategies}\label{ssec:method-eviction}
\begin{figure}[!ht]
    \centering
    \includegraphics[width=\linewidth]{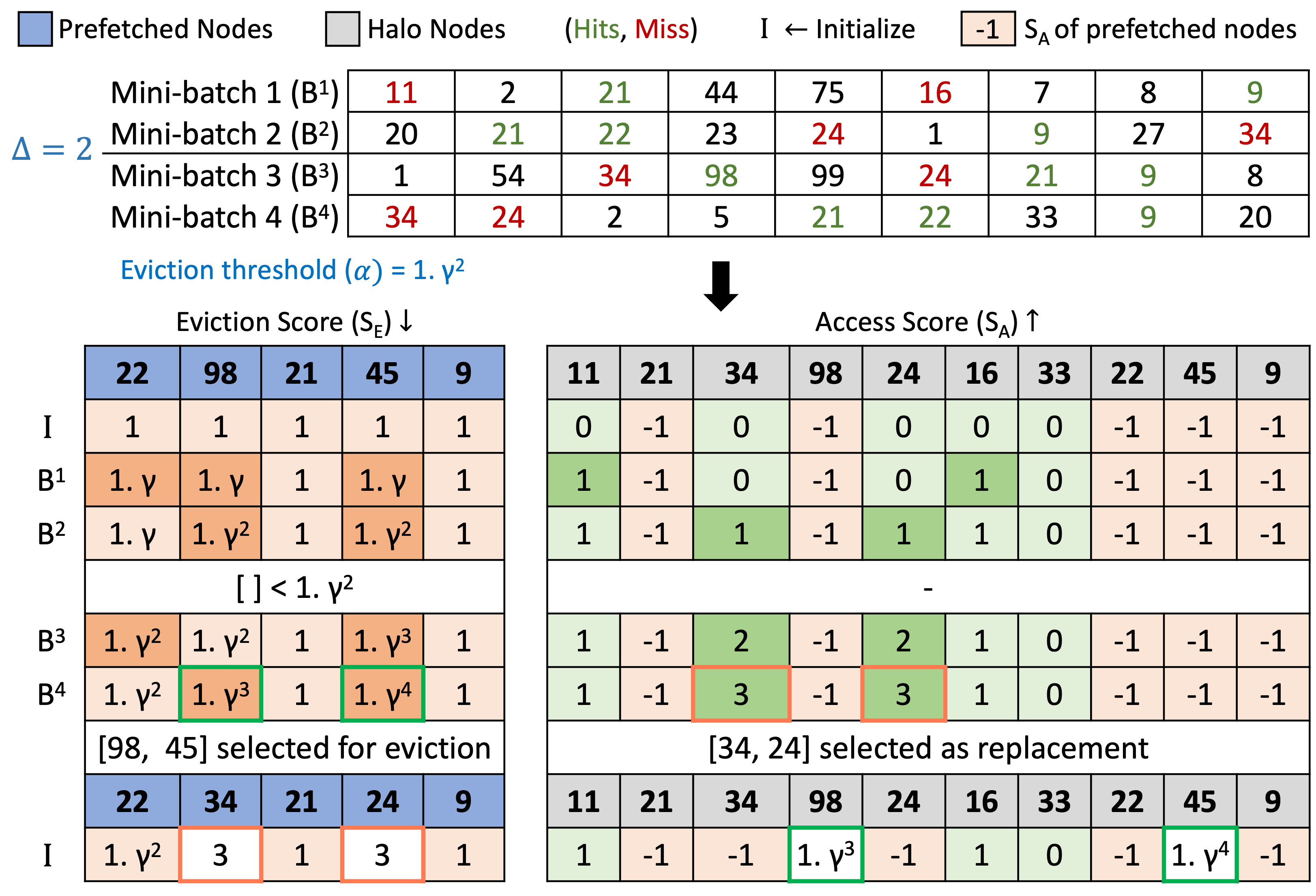}
    \caption{Visual representation of $S_E$ and $S_A$ scoreboards showing how prefetched nodes are selected for eviction and new candidates are chosen for replacing them. Highlighted borders represent swapping.}
    \label{fig:scores}
\end{figure}
To complement the prefetch buffer, we introduce a dynamic score-based eviction strategy (Fig. \ref{fig:scores}). We use two distinct scoring mechanisms: \emph{Eviction score} ($S_E$) and \emph{Access score} ($S_A$). $S_E$ is used to determine the least utilized nodes (Algorithm~\ref{algo:prefetch-evict}, line \#29: $V_{p}^{h|e}$) in the prefetch buffer for potential eviction. On the other hand, $S_A$ identifies the most suitable replacement candidates (Algorithm~\ref{algo:prefetch-evict}, line \#31: $V_{p}^{r|e}$) by scoring the nodes that were frequently sampled but missed in the prefetch buffer. Both scores are continuously updated in every minibatch and the underlying computation is hidden by the RPC calls fetching the missed halo nodes sampled in the minibatch. $S_A$ for all prefetched nodes is set to $-1$, while for other remote nodes of the partition, it is set to 0. Whenever a node is sampled during training but is found to be missing in the prefetch buffer (a `miss'), its corresponding $S_A$ is incremented by 1. As training progresses, nodes that are frequently sampled but were not prefetched accumulate higher $S_A$, signaling their increased likelihood of being needed in the future. 

$S_E$ is initialized to 1 for the nodes present in the prefetch buffer (all nodes are initially equally likely to be evicted) and decays over time. When a node in the prefetch buffer is not used during a training iteration, its $S_E$ is multiplied by the decay factor $\gamma$. A lower $\gamma$ value results in a high decay rate, indicating a higher likelihood of eviction for nodes that are not recently used. In contrast, a high $\gamma$ is equivalent to low decay. The eviction process is triggered at regular intervals of $\Delta$. During an eviction round (line \#13 in Algorithm~\ref{algo:prefetch-evict}), $S_E$ of the prefetched nodes are assessed. Nodes whose eviction scores have dropped below a predefined threshold $\alpha$ are selected for eviction. If $k$ nodes are identified as candidates for eviction, we then select the highest degree (top-$k$) nodes with the highest $S_A$ as their replacements. After successful eviction and replacement, we initialize the evicted nodes' new $S_A$ (previously $-1$) to their last $S_E$ and the replacements' initial $S_E$ to their last $S_A$ (swapping). The number of nodes chosen for replacement is exactly equal to the number of nodes evicted to maintain a constant buffer size. The threshold is defined in Equation~\ref{eqn:alphathreshold}. 
\begin{equation}\label{eqn:alphathreshold}
{\alpha} = S_E \times \gamma^{\Delta}
\end{equation}
Our standard $S_A$ is a \texttt{numpy} array of length $\mathcal{O}(|V|)$ (Table~\ref{tab:notation}) which reduces the indexing overhead to $\mathcal{O}(1)$ during repetitive updates of $S_A$ on every minibatch. However, since the memory consumption of $\mathcal{O}(|V|)$ is not suitable for large graphs, our implementation also supports a \textbf{\emph{memory-efficient}} version of $S_A$ where we reduce the size to $\mathcal{O}(|V_{p}^{h}|)$. Reduction in memory enhances the cost of searching by $\mathcal{O}(log(V_{p}^{h}))$; we perform binary search to locate and update $S_A$ in parallel. 

\subsection{Performance Model}\label{ssec:method-model}
We define the overall time (denoted by $T$) in terms of the individual times (depicted by $t$) spent by the constituent workloads per minibatch. For baseline DistDGL (denoted as $T_{Baseline}$), a significant amount of the time per minibatch is spent in sampling (denoted by $t_{Sampling}$) neighboring nodes (neighborhood size is determined by specified fan-out per hop of the GNN layer), fetching node features across the partitions (represented by the maximum of remote RPC communication and local copy from the current partition, $\mathrm{max}(t_{RPC}, t_{Copy})$) and finally invoking data parallel training on the distributed graph (Distributed Data Parallel (DDP) training time is referred by $t_{DDP}$). Equation~\ref{eqn:baseline} depicts the baseline time.
\begin{multline}\label{eqn:baseline}
T_{Baseline} = \\ \underbrace{t_{Sampling}}_\text{Nodes for gathering msgs.} + \underbrace{\mathrm{max}(t_{RPC}, t_{Copy})}_\text{Fetch local\slash  remote features} + \underbrace{t_{DDP}}_\text{Data Parallel Training}
\end{multline}
\normalsize

We undertake an intermediate step of concurrently preparing the next minibatch through the prefetching scheme, overlapped with training on the current minibatch. Consider the minibatch preparation time for an arbitrary minibatch $s$ as $t_{Prepare}$, which constitutes neighbor sampling ($t_{Sampling}$), inspecting candidate nodes in the prefetch buffer to determine eviction or replacement ($t_{Lookup}$), maintaining scoreboard to determine candidate node longevity in the prefetch buffer ($t_{Scoring}$) and data movement to assemble requisite node features (similar to baseline, represented by the maximum of remote communication and local copy, $\mathrm{max}(t_{RPC}, t_{Copy}$). Derivation of $t_{Prepare}$ is shown in Equation~\ref{eqn:t-prepare}.
\begin{multline}\label{eqn:t-prepare}
t_{Prepare} = \underbrace{t_{Sampling}}_\text{Nodes for gathering msgs.} + \underbrace{t_{Lookup}}_\text{Inspect prefetched nodes} + \\
    \underbrace{\mathrm{max}\left(\underbrace{t_{Scoring}}_\text{Maintain scoreboard}, \underbrace{\mathrm{max}(t_{RPC}, t_{Copy})}_\text{From $T_{Baseline}$}\right)}_\text{Overlap fetching features with score update}
\end{multline}
\normalsize

There is a distinction between the first vs. rest of the minibatches, in terms of the possible overlap that can be achieved by engaging concurrent resources for training and minibatch preparation. The first minibatch must invoke $t_{Prepare}$ in addition to performing training and preparing for the next minibatch concurrently (as shown in Equation~\ref{eqn:t-mini1}), whereas rest of the minibatches can leverage the prefetching initiated on the previous minibatch, and achieve a greater overlap between DDP training on the current minibatch and the next minibatch preparation (represented by Equation~\ref{eqn:t-minin}).  
\begin{equation}\label{eqn:t-mini1}
T^{s{[0]}}_{Prefetch}= \underbrace{t_{Prepare}}_\text{For \textit{current} minibatch} + \underbrace{\mathrm{max}(t_{Prepare}, t_{DDP})}_\text{Overlap  training and \textit{next} minibatch} 
\end{equation}
\normalsize
\begin{equation}\label{eqn:t-minin}
T^{s{[1:]}}_{Prefetch}= \underbrace{\mathrm{max}(t_{Prepare}, t_{DDP})}_\text{Training on \textit{prefetched} minibatch overlapped with next minibatch} 
\end{equation}
\normalsize

We anticipate a \emph{perfect overlap} scenario where $t_{Prepare}\leq t_{DDP}$, leading to $T_{Prefetch}<T_{Baseline}$ (required for overall performance improvement). We observe this \emph{perfect overlap} often in CPU-based training (see \S\ref{sec:eval}), as it takes relatively longer than on GPUs, making it more likely to achieve an overlap with the minibatch preparation. On the other hand, if $t_{Prepare} \geq t_{DDP}$, then there is either no overall improvement or some improvement when the sampling time is predominant, i.e., $t_{Prepare}-t_{DDP} < t_{Sampling}$. In practice, Equation~\ref{eqn:t-minin} dominates the time (across hundreds of minibatches), so the contribution of Equation~\ref{eqn:t-mini1} becomes inconsequential. Thus, $T_{Prefetch} \equiv T^{s{[1:]}}_{Prefetch} \approx t_{DDP}$. Dividing $T_{Baseline}$ (i.e., Equation~\ref{eqn:baseline}) with $T_{Prefetch}$ gives the potential improvement factor, as shown in Equation~\ref{eqn:prebase}. Sampling leads to feature data movement, for simplicity we assume $t_{Sampling}$ to be less expensive relative to the communication overhead, $t_{RPC}$.
\begin{equation}\label{eqn:prebase}
\frac{t_{Sampling} + \mathrm{max}(t_{RPC}, t_{Copy})}{t_{DDP}} + 1 
\approx \frac{t_{RPC}}{t_{DDP}} + 1
\end{equation}
\normalsize
Equation~\ref{eqn:prebase} indicates that if $\frac{t_{RPC}}{t_{DDP}}$ is greater than $1$ (feature data movement exceeding training expenses), then the proposed prefetching scheme can lead to major performance improvements (considering a decent hit rate). This situation is more likely, since graph datasets are multifarious and graph neighborhoods in contemporary GNN models demonstrate significant expansion possibilities. Hence, distributed GNN training platforms must contend with the massive increase in the data movement prior to data-parallel training. However, if communication is not on the critical path, proposed approach may not be effective in achieving perfect overlap. 

Another aspect preventing a potential overlap is when the scoring overheads in $t_{Prepare}$ becomes non-trivial due to relatively frequent score maintenance as predicated by the choice of the interval and decay factor, i.e., $\Delta$ and $\gamma$ ($\Delta$ is more relevant here), leading to a compounding effect. Let us consider a fixed $t_{Scoring}$ per interval and formulate the cumulative effect of overheads by following the \emph{compound interest} formula where $t_{Scoring}$ represents the \emph{interest}, $t_{Prepare}$ is the \emph{principal} and \emph{time period} ($t$) corresponds to the number of intervals of score maintenance (i.e., $t=\frac{\#epochs}{\Delta}$) :
     \begin{equation}\label{eqn:scoring-overhead}
     t_{Prepare} (future) = t_{Prepare} (present) \times (1 + \frac{t_{Scoring}}{100})^{t}
     \end{equation}
Considering $t_{Scoring}$ to be around 10\% of $t_{Prepare}$ and $t = 10$ (\#epochs=100 and $\Delta=10$), upon substituting the values in Equation~\ref{eqn:scoring-overhead}: $t_{Prepare}(future) = t_{Prepare}(present) \times (1.1)^{10}$ (implying $\sim25\%$ overhead across 100 epochs for maintaining the score once in every 10 epochs). Therefore, lower values of $\Delta$ (i.e., frequent scoreboard maintenance) can lead to significant overheads, leading to $t_{Prepare} > t_{DDP}$ (impact of score maintenance is discussed in \S\ref{sssec:analysis-gamma}).

\subsection{Metrics}\label{ssec:method-metrics}
We use \emph{hit rate} as a metric to evaluate our prefetch and eviction scheme. In the context of distributed GNN training, $h$ is the number of remote nodes sampled and were found to be ``prefetched'' in the buffer, and $m$ is the number of sampled remote nodes not present in the buffer, the \emph{hit rate} is:
\begin{equation}
\text{hit rate} = \frac{h}{h+m}
\end{equation}

We empirically observed that an increase in hit rate may not correspond to performance improvements (refer to \S\ref{sssec:eval-baseline-hitrate}). This discrepancy primarily arises from two factors: (1) performance improvement is not only governed by a decrease in communication overhead but also by the effective overlap of training and sampling\slash lookup for next minibatch preparation; (2) If the local access time for the prefetched nodes exceeds the time spent on remote communication for fetching these nodes ($t_{Copy} > t_{RPC}$ in \S\ref{ssec:method-model}), then a higher hit rate may not lead to an overall improvement in performance.

\subsection{Prefetch and Eviction Trade-offs}\label{ssec:method-tradeoff}
\begin{figure}[!ht]
    \centering \includegraphics[width=0.9\linewidth]{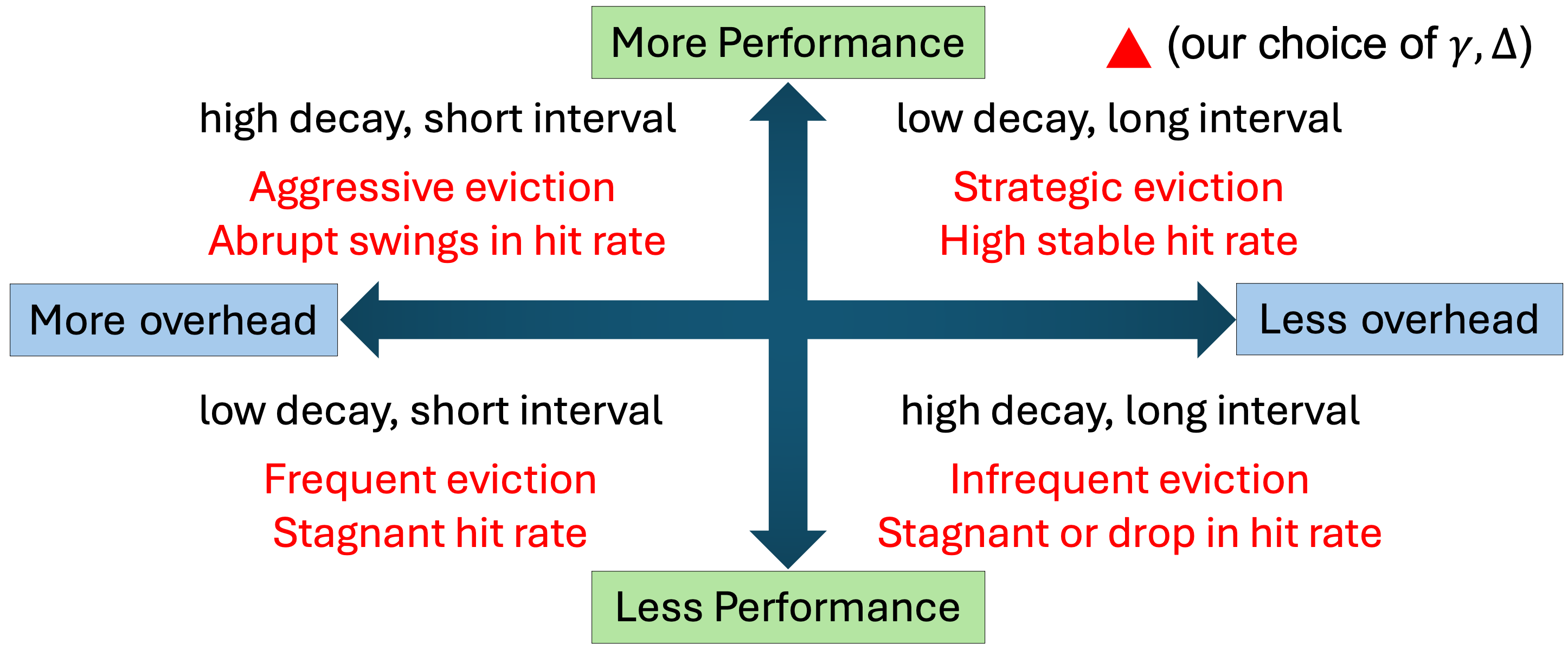}
    \caption{Quadrants showing four different tradeoff scenarios using various combinations of decay factor ($\gamma$) and eviction interval ($\Delta$).}
    \label{fig:tradeoff}
\end{figure}
We present four scenarios in our prefetch and eviction scheme highlighting the trade-offs between high and low eviction intervals and the decay of eviction scores. The four quadrants as shown in Fig. \ref{fig:tradeoff} illustrate the following scenarios.
\begin{enumerate}[wide, labelwidth=!, labelindent=0pt]
 \item \emph{Low decay, short interval}: When the decay is low ($\gamma$ close to $1$), the buffer's eviction scores are gradually reduced. A short interval results in frequent eviction rounds. When both of these conditions are combined, the hit rate can exhibit stagnation due to high tolerance for unused nodes on the prefetch buffer (fewer nodes evicted per round). Additionally, with relatively frequent eviction rounds, computational overhead in inspecting which nodes can be evicted and replaced with better candidates can be significant.
 \item \emph{High decay, short interval}: With high decay (i.e., $\gamma$ close to $0$) and short intervals, scores are reduced more aggressively. In addition, frequent evictions occur due to low intervals, leading to an upsurge in evictions. This risks evicting useful nodes, making the scheme less tolerant and leading to swings in the hit rate across minibatches. Short interval also entails more overhead due to repeated inspection of eviction and replacement candidates.
 \item \emph{High decay, long interval}: When both decay and interval are high, the eviction scores are reduced over a relatively longer period which results in delayed evictions and fewer replacements. This is likely to result in less performance and possible hit rate stagnation like the low decay, low interval scenario, with chances of drops in hit rates due to the longer intervals. With long intervals, the overhead remains low.
 \item \emph{Low decay, long interval}: This scenario represents strategic eviction with the goal of consistent hit rate growth, exhibiting better performance with reduced overhead. Nodes are not evicted aggressively, rather, the low decay (i.e., $\gamma$ close to $1$) gradually diminishes the eviction scores of the prefetched nodes over a relatively longer interval, promoting efficient buffer utilization, demonstrating appropriate trade-off.
\end{enumerate}
\begin{figure*}[!t]
    \centering
        \begin{subfigure}{0.9\textwidth}
            \includegraphics[width=\linewidth]{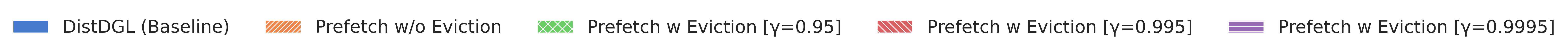}
            \label{fig:legend}
        \end{subfigure}
        \begin{subfigure}{0.24\textwidth}
        \includegraphics[width=\linewidth]{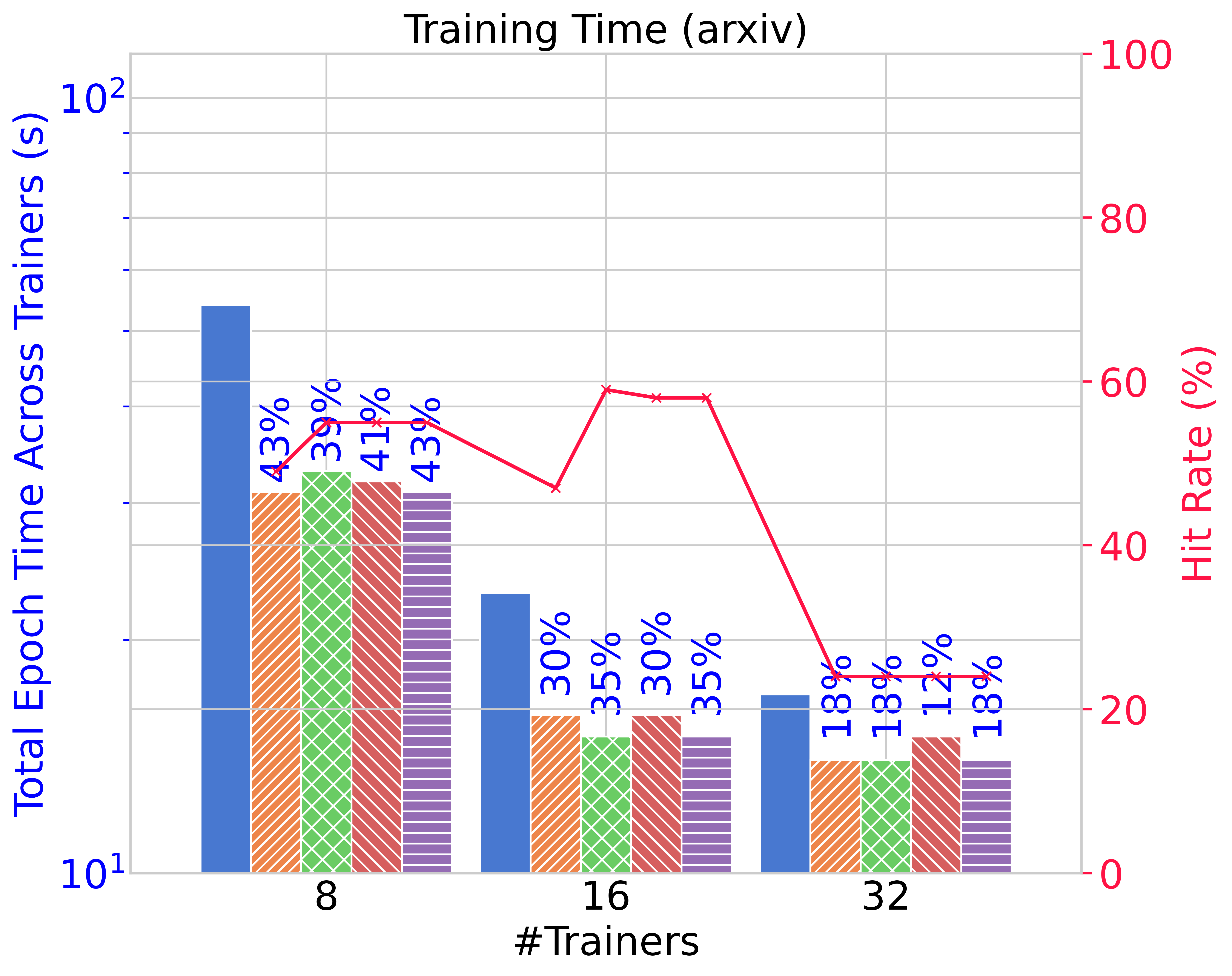}
        \caption{Arxiv-CPU}
        \label{fig:arxiv-cpu}
    \end{subfigure}
    \begin{subfigure}{0.24\textwidth}
        \includegraphics[width=\linewidth]{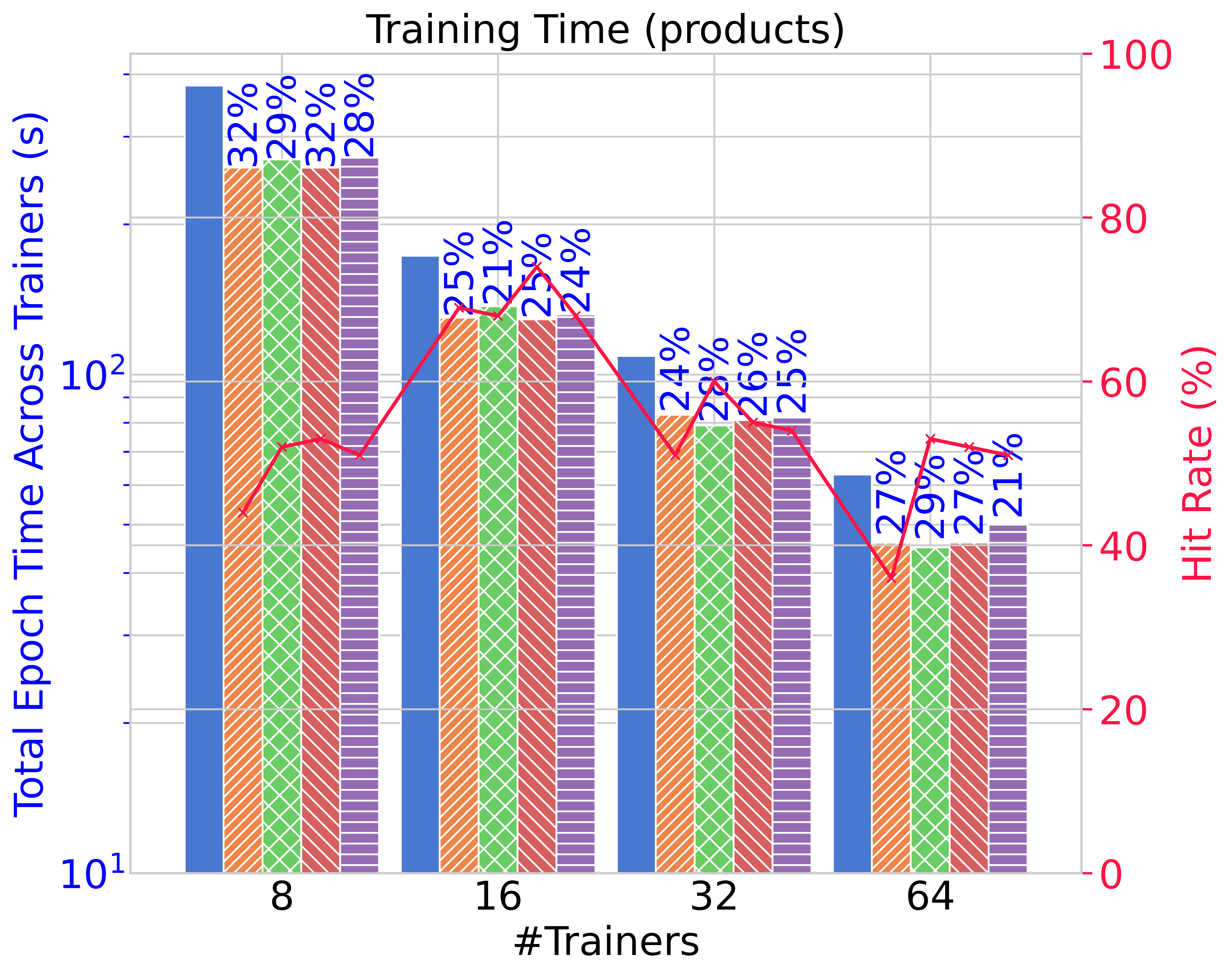}
        \caption{Products-CPU}
        \label{fig:products-cpu}
    \end{subfigure}
    \begin{subfigure}{0.24\textwidth}
        \includegraphics[width=\linewidth]{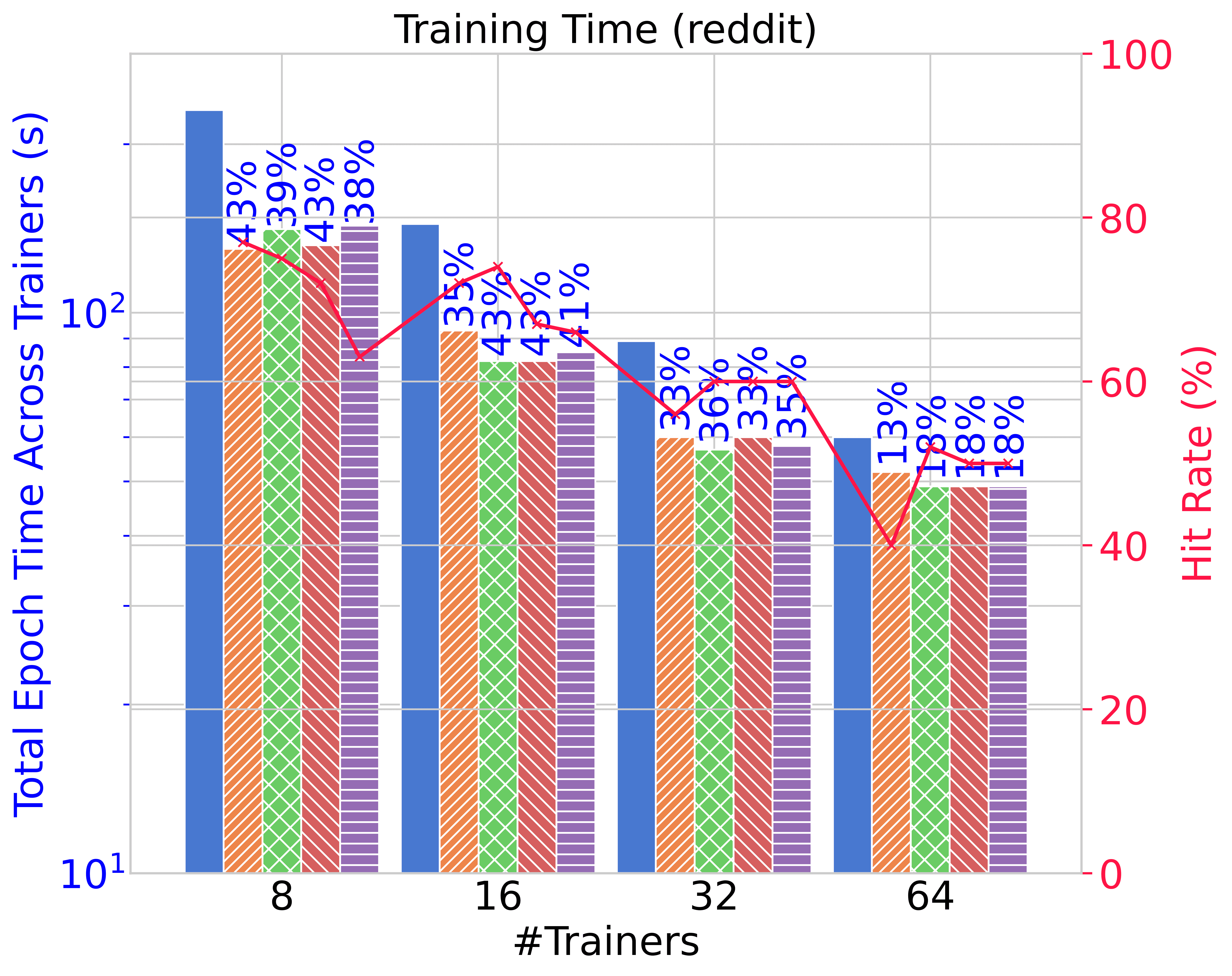}
        \caption{Reddit-CPU}
        \label{fig:reddit-cpu}
    \end{subfigure}
    \begin{subfigure}{0.24\textwidth}
        \includegraphics[width=\linewidth]{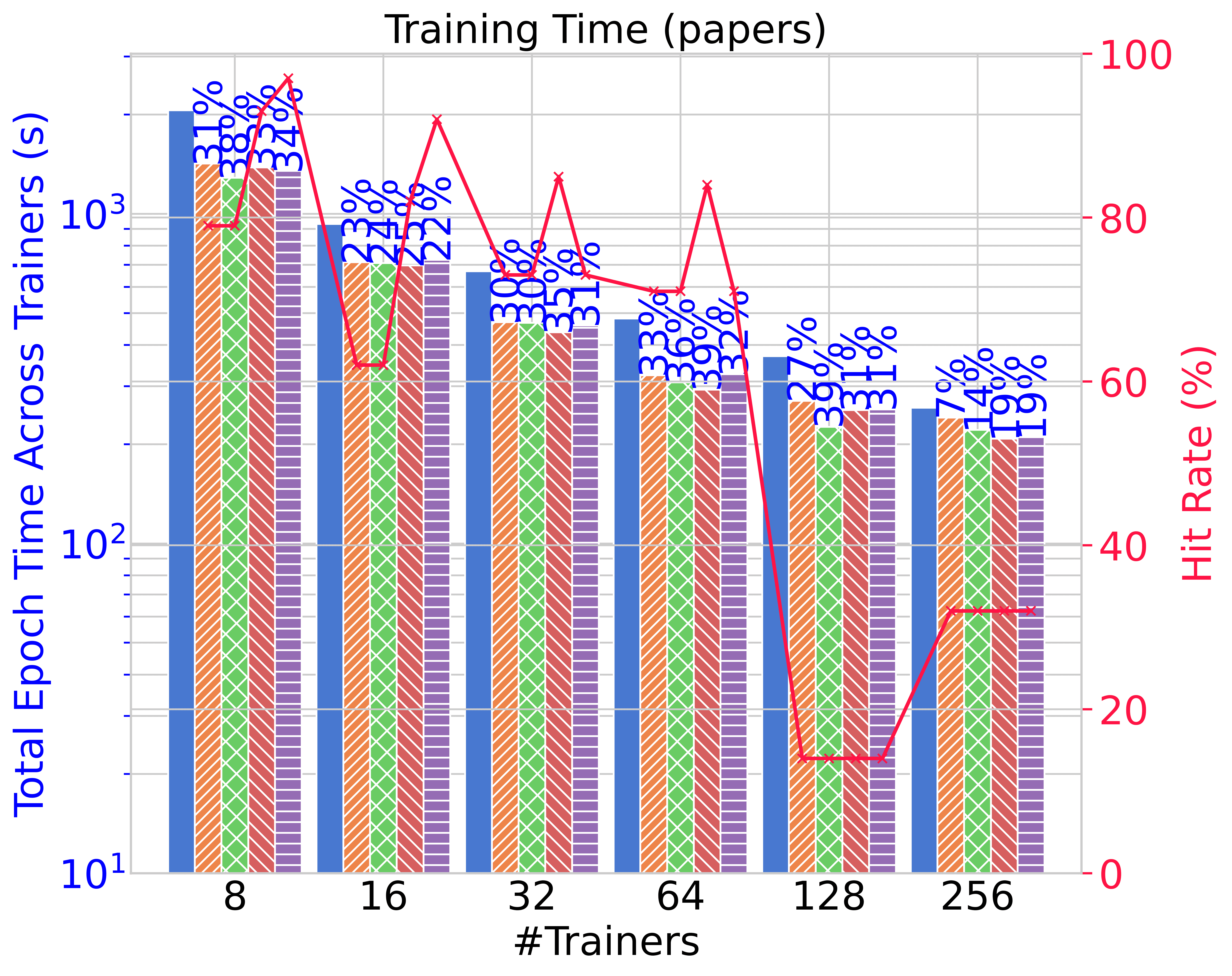}
        \caption{Papers-CPU}
        \label{fig:papers-cpu}
    \end{subfigure}
    \begin{subfigure}{0.24\textwidth}
        \includegraphics[width=\linewidth]{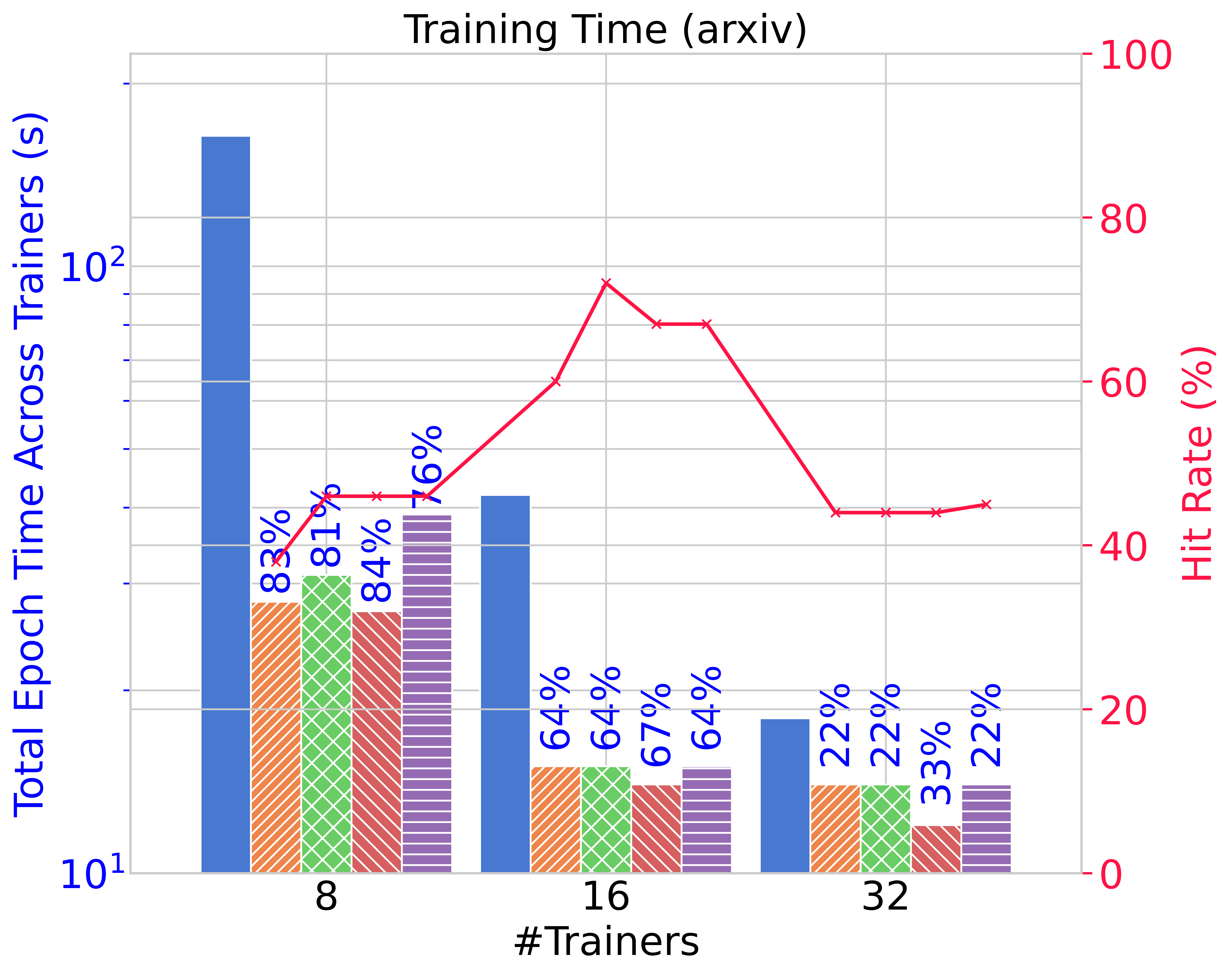}
        \caption{Arxiv-GPU}
        \label{fig:arxiv-gpu}
    \end{subfigure}
    \begin{subfigure}{0.24\textwidth}
        \includegraphics[width=\linewidth]{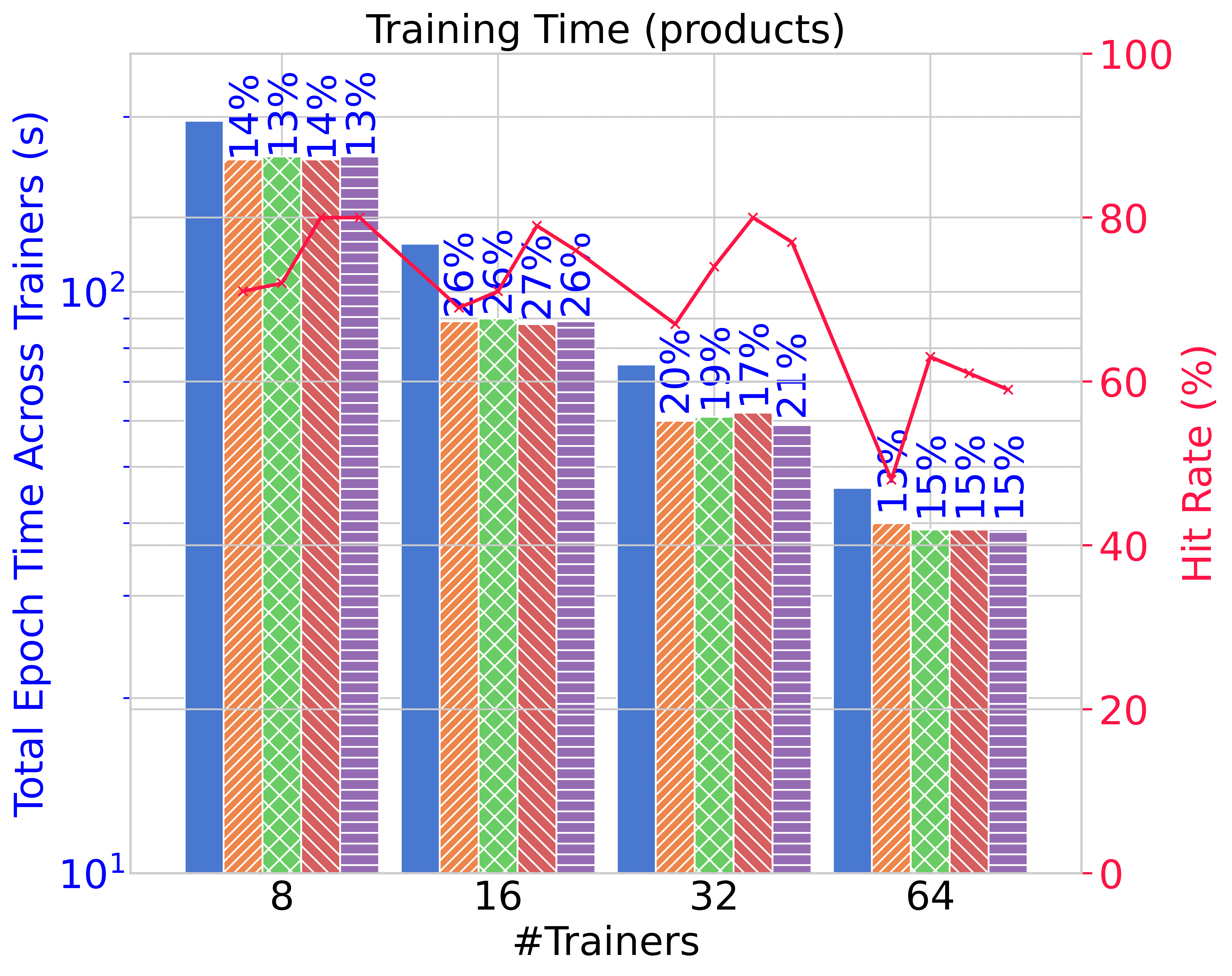}
        \caption{Products-GPU}
        \label{fig:products-gpu}
    \end{subfigure}
    \begin{subfigure}{0.24\textwidth}
        \includegraphics[width=\linewidth]{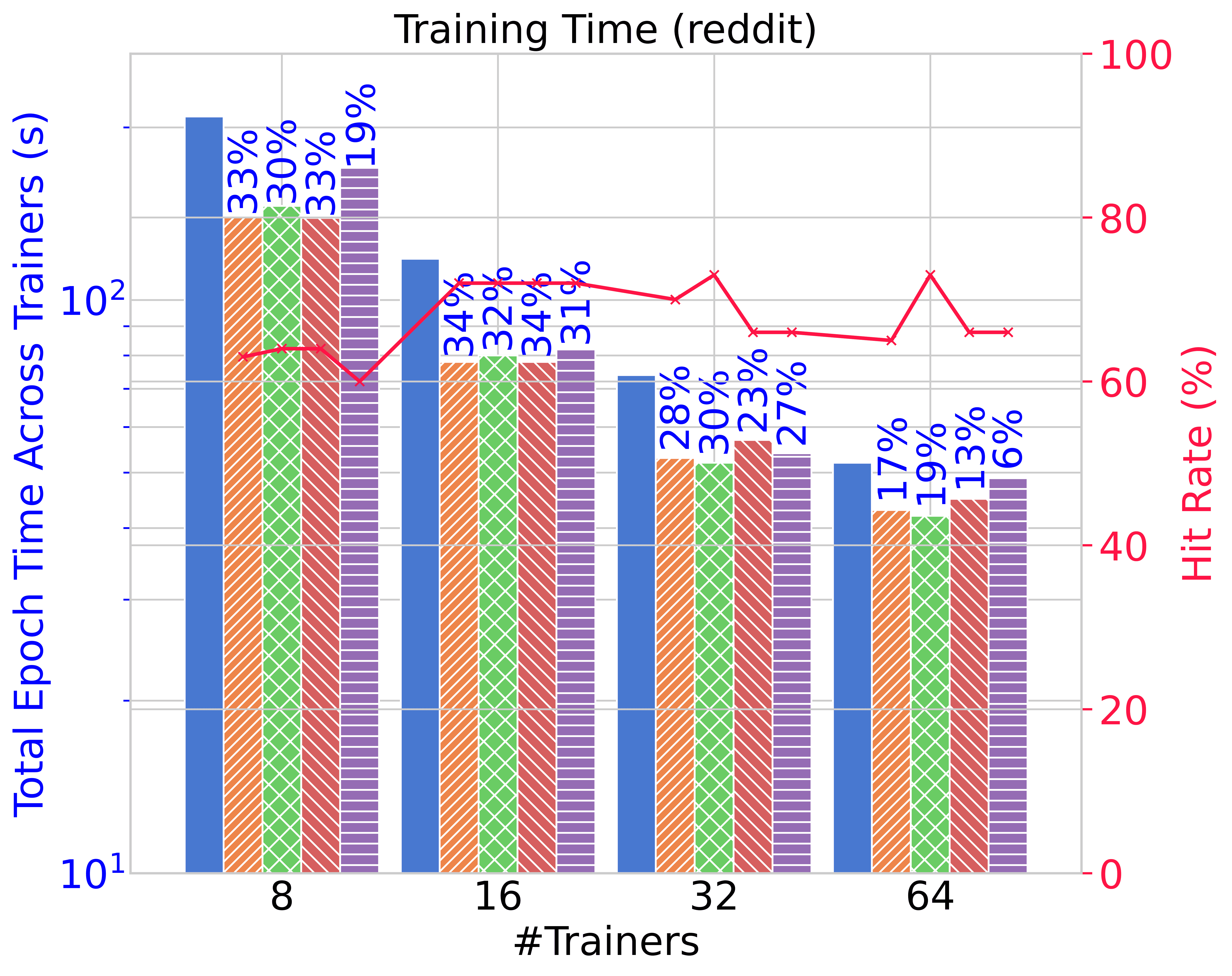}
        \caption{Reddit-GPU}
        \label{fig:reddit-gpu}
    \end{subfigure}
    \begin{subfigure}{0.24\textwidth}
        \includegraphics[width=\linewidth]{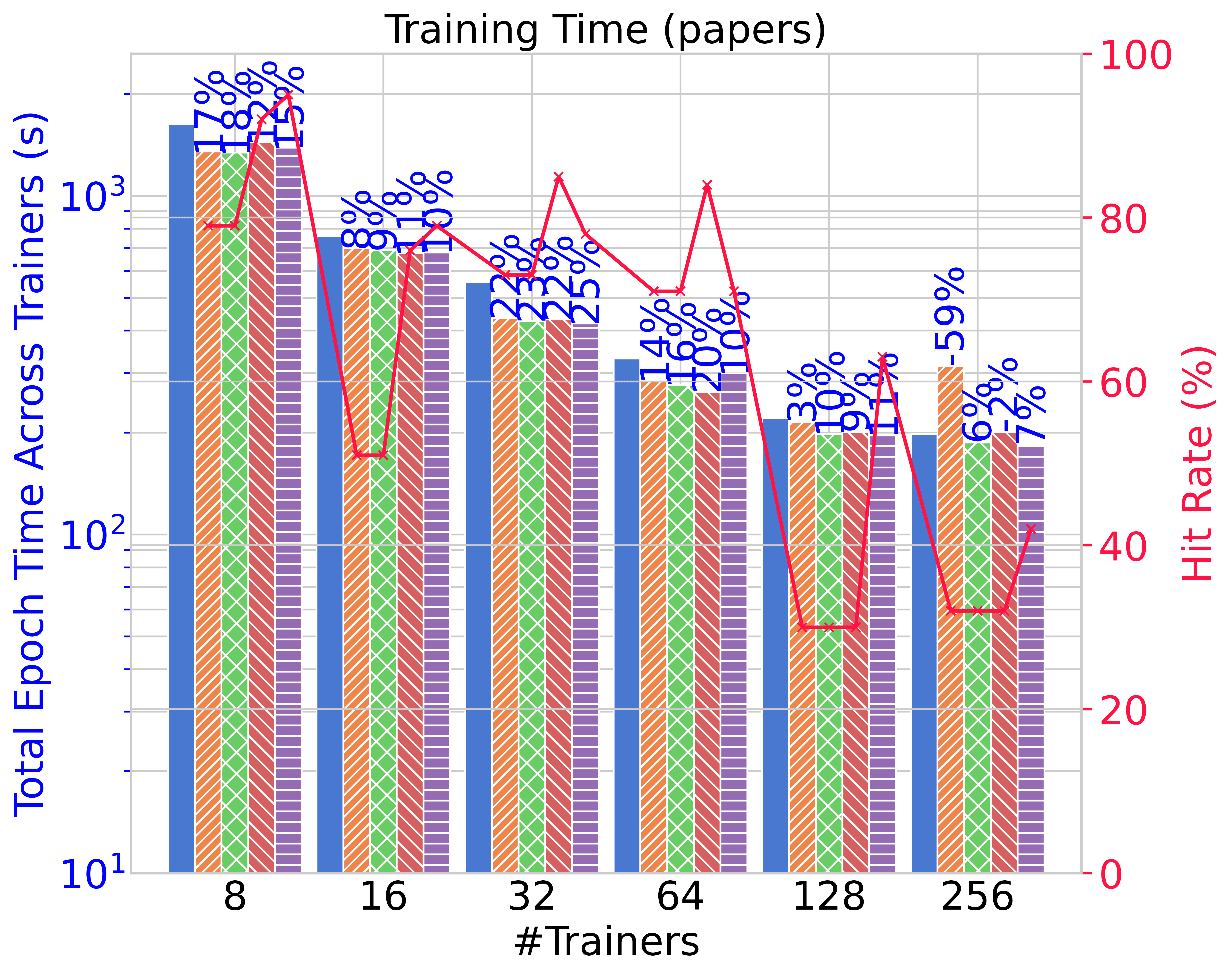}
        \caption{Papers-GPU}
        \label{fig:papers-gpu}
    \end{subfigure}
    \caption{Training performance of our prefetching and eviction scheme on GraphSAGE against DistDGL across 4 inputs, 2 versions (CPU, GPU), and up to 64 node configurations (4 trainers/node). In each figure, the annotations on top of each bar highlight the percent reduction in the execution times in MassiveGNN vs. DistDGL. The optimal $f_{p}^{h}$, $\gamma$ and $\Delta$ yielding maximum improvement are listed in Table \ref{tab:cpu_gpu_params}. [Y1: Lower is better; Y2: Higher is better]}
    \label{fig:baseline}
\end{figure*}

\section{Evaluation and Analysis}\label{sec:eval}
The experiments are performed on the NERSC Perlmutter supercomputer consisting of 1,792 GPU, and 3,072 CPU-only compute nodes. Each node uses 64-core 2.4GHz AMD EPYC 7763 CPUs with 256GB of DDR4 memory, 256MB L3 cache, and 8 memory channels (GPU nodes have 4$\times$ NVIDIA A100 GPUs, we use nodes with 40GB GPU memory) with HPE Slingshot 11 interconnect~\cite{yang2020accelerate}. In this section, after briefly discussing the experimental configurations and data, we examine CPU\slash GPU training performances in \S\ref{ssec:eval-baseline} and provide detailed analysis in \S\ref{ssec:analysis}. 
\paragraph*{Utilizing CPU resources}\label{sec-par:eval-resources}
We use Python \texttt{ThreadPoolExecutor}\footnote{\url{https://docs.python.org/3/library/concurrent.futures.html\#concurrent.futures.ThreadPoolExecutor}} to enable concurrency on I/O-bound tasks, configured with workers equal to the desired number of look-ahead minibatches (currently set to 1). The cost of forking threads is incurred once, as the same threads are reused throughout training. We engage CPU cores (which would otherwise be left idle) in preparing future minibatches. 
\texttt{ThreadPoolExecutor} is not able to circumvent Python GIL, which constraints multithreading; we use NUMBA~\cite{lam2015numba} Just-In-Time (JIT) features to enable multithreading using OpenMP bypassing the Python APIs, and releasing the GIL.
For CPU training, we use 32 cores\slash trainer; 16 cores are dedicated to NUMBA~\cite{lam2015numba} parallel operations, and the remaining 16 are allocated for PyTorch DDP. Threads for both NUMBA and PyTorch are bound locally to their respective cores, preventing thread migration across sockets. 
\paragraph*{Dataset, Model and Prefetching Parameters}\label{sec-par:eval-datasets}
We conducted our experiments using four datasets listed in Table~\ref{tab:datasets}. We focus on the node classification task in homogeneous graphs and primarily use a 2-layer GraphSAGE \cite{hamilton2017inductive} with a fanout of \{10, 25\} and a batch size of $2000$. We evaluate our prefetching scheme in terms of training time and hit rate for 2--64 nodes (4 GPUs\slash node). The GNN's accuracy remains unchanged from the baseline version because our prefetching scheme optimizes the pre-training data pipeline without altering the underlying training process or model architecture. In all our experiments, each node has four trainer processes and runs training for 100 epochs.  
\begin{table}[!ht]
\caption{Datasets used in the experiments.}
\centering
\begin{tabular}{|l||r|r|r|}
\hline
\textbf{Dataset}       & \textbf{Nodes $\mathbf{|V|}$} & \textbf{Edges $\mathbf{|E|}$}     & \textbf{Feature Dimension} \\ \hline \hline
arxiv~\cite{hu2020open}                 & 0.16M       & 1.16M	     & 128\\ \hline
products~\cite{hu2020open}         & 2.4M        & 61.85M        & 100\\ \hline
reddit~\cite{9407264}              & 0.23M          & 114.61M        & 602\\ \hline
papers~\cite{hu2020open}              & 111M        & 1.6B         & 128\\ \hline
\end{tabular} 
\label{tab:datasets}
\end{table}
The number of graph partitions is equal to the number of nodes used in an experiment. We test various values of $f_{p}^{h}$ (15\%, 25\%, 35\%, and 50\%), $\Delta$ (16, 32, 64, 128, 512, and 1024) and $\gamma$ (0.95, 0.995, 0.9995). The choice of $\gamma$ was found empirically and discussed in \S\ref{sssec:analysis-gamma}.
\paragraph*{DistDGL (Baseline)}\label{sec-par:eval-dgl-settings} We used DistDGL v1.1.3, PyTorch v1.13 with NCCL v2.14.3 and CUDA v11.7. For the CPU training, we use PyTorch Gloo\cite{gloo} backend as recommended. We configure DistDGL with 1 server per node, and the number of sampler processes is set to 0. DGL's partitioning API partitions the graph datasets using METIS~\cite{karypis1998fast}. We uniformly sample a fixed-size set of local node neighbors using a \texttt{NeighborSampler}~\cite{hamilton2017inductive}, which applies a fan-out strategy to keep subset sizes consistent across layers for training the GNN model. This fan-out method involves selecting a predetermined number of neighbors for each node, thereby controlling the expansion of the computation graph and ensuring manageable complexity while preserving essential neighborhood structure information. Our approach is compatible with other sampling methods as well.
The code is derived from DistDGL (v1.1.3) and available on GitHub with a permissive license: \url{https://github.com/pnnl/MassiveGNN}.

\subsection{Overall training performances}\label{ssec:eval-baseline}
The proposed prefetch method supports both CPU and GPU backends during training; we examine the trade-offs associated with two configurations: \emph{prefetch without eviction} and \emph{prefetch with eviction}. \emph{Prefetch without eviction} represents the version where remote nodes of the partitions are prefetched by each trainer only once during buffer initialization and retained without eviction throughout the training (line \#2 in Algorithm~\ref{algo:gnn-training}). We test against various buffer sizes ($f_{p}^{h}$ denotes the fraction of halo nodes; see Table~\ref{tab:notation}). On the other hand, \emph{prefetch with eviction} represents periodic eviction of nodes from the buffer (see Algorithm~\ref{algo:prefetch-evict}). We evaluate this version with various combinations of $ \Delta$ and $\gamma$. Fig.~\ref{fig:baseline} shows the best improvements on GPU and CPU (with different optimal $f_{p}^{h},\Delta, \gamma$) against the datasets. Overall, we observe about 15--40\% improvements (the highest was about 85\% for arxiv) relative to baseline DistDGL. 

To keep the wall clock time within bounds for distributed workloads, we have chosen a constant batch size of $2000$ across all configurations (like ~\cite{zheng2020distdgl}). As a result, as the \#trainers increase, leading to more \#partitions being trained on, each trainer owns fewer nodes of the graph. With a constant batchsize this reduces the overall \#minibatches each trainer processes as shown in Table~\ref{tab:num_halo}. We discuss GPU\slash CPU results, hit rate trends, and scalability in the forthcoming sections. 

\begin{table}[h]
\caption{Average number of remote nodes per trainer across datasets (left) paired with the total number of minibatches each trainer processes (right), with a constant batch size of 2000 over $100$ epochs.}
\centering
\begin{tabular}{|l||r|r|r|r|}
\hline
\textbf{\#Trainers}       & \textbf{arxiv} & \textbf{reddit} & \textbf{products}   & \textbf{papers} \\ 
\hline \hline
8 & 34.6K/600& 91.5K/1000 & 471K/1300 & 14.9M/7600 \\
\hline
16 & 30.6K/300 & 119.7K/500 & 352.3K/700 & 14.8M/3800 \\
\hline
32 & 26.2K/200 & 123.9K/300 & 307.8K/400 & 14.1M/1900\\
\hline
64 & - & 120.7K/200  & 220.5K/200 & 10.4M/1000\\
\hline
128 & - & - & - & 7.7M/500 \\
\hline
256 & - & - & - & 4.8M/300\\
\hline
\end{tabular} 
\label{tab:num_halo}
\end{table}

\subsubsection{CPU results}\label{sssec:eval-baseline-cpu} 
\paragraph*{Prefetch without Eviction} In Fig. \ref{fig:baseline} [a-d], the optimal $f_{p}^{h}$ (indicated by the orange bar in each group) consistently demonstrates improvements through our degree-based prefetching strategy across various datasets and node configurations.  When training occurs exclusively on CPUs, as depicted in Fig. \ref{fig:baseline} [a-d], we see significant performance improvements over DistDGL, up to a maximum of $32-43\%$ across datasets. The slower nature of CPU training allows for a \emph{perfect} overlap between the training of current minibatches and the preparation of subsequent ones. All configurations tested underscore the advantages of degree-based prefetching, consistently demonstrating improvements across the spectrum. 

\paragraph*{Prefetch with Eviction} We explore our eviction strategy by optimizing $\Delta$ over optimal $f_{p}^{h}$ and assessing the impact of our chosen $\gamma$ values. In Fig. \ref{fig:baseline} [a-d], the 3 bars from the right in each group depict different $\gamma$ over optimal $f_{p}^{h}$ and $\Delta$. We observe that the eviction strategy further improves performance on CPU by $5-12$ percent points, which shows the efficiency of the proposed score-based strategy.

\subsubsection{GPU results}\label{sssec:eval-baseline-gpu}

\paragraph*{Prefetch without Eviction} On GPU training (Fig. \ref{fig:baseline} [e-h]), we observe maximum improvements of $26\%-34\%$ across datasets and specifically about $85\%$ in arxiv. We observe a performance degradation in papers on 64-nodes, which is a side effect of subpar hit rate and relatively fewer minibatches (see Table~\ref{tab:num_halo}), pushing the local access overheads to exceed communication prior to training, preventing a \emph{perfect overlap} (see \S\ref{ssec:method-metrics}). The situation improves as hit rates rebound when eviction is enabled.

\paragraph*{Prefetch with Eviction}  Adding an eviction strategy partially addresses this degradation and shows a further increase in performance by $1-8$ percent points across datasets (smaller compared to CPUs). In all the experiments, papers use memory-efficient eviction and hence have a slightly higher computational overhead than other inputs. As discussed before, papers on 64-nodes exhibit second-order effects of lower hit rates and fewer minibatches, and despite eviction trade-offs (\S\ref{ssec:method-tradeoff}), we observe about 7\% improved performance as hit rates rebound for highest $\gamma$. Arxiv exhibits a relatively large diameter and small degree \cite{hu2020open}, resulting in severe load imbalance (spends 6x more time on communication and data movement than training, exacerbated by relatively small size) that negates the computational advantages on GPUs.

\subsubsection{Hit Rate}\label{sssec:eval-baseline-hitrate} 
Fig. \ref{fig:baseline} also shows hit rate growth across all datasets and in almost all configurations. We observe that the hit rate decreases as the number of trainers increases, which corresponds to a reduction in partition size per trainer. Since the total number of minibatches decreases with an increase in number of trainers, as seen in Table \ref{tab:num_halo}, due to constant batch sizes, each trainer's buffer receives fewer minibatches to optimize on.  However, it is evident from our experiments on a higher number of minibatches (later discussed in \S\ref{sssec:analysis-eviction}) that hit rates do increase as trainers get sufficient number of minibatches. We also observe that although hit rates do not always linearly correlate with performance (as discussed in \S\ref{sssec:analysis-metric}), comparisons between scenarios with and without eviction consistently show that the hit rate remains flat when eviction does not further enhance performance (Fig. \ref{fig:baseline}g:16 trainers, Fig. \ref{fig:baseline}:32 trainers).

\begin{table*}[!ht]
\caption{Optimal combination of $f_{p}^{h}$, $\gamma$ and $\Delta$ for GraphSAGE across datasets for both CPU and GPU versions.}
\centering
\resizebox{\textwidth}{!}{
\begin{tabular}{|c||c|c|c|c||c|c|c|c|}
\hline
\textbf{\#Nodes} & \multicolumn{4}{c||}{\textbf{CPU Version ($f_{p}^{h}$, $\gamma$, $\Delta$)}} & \multicolumn{4}{c|}{\textbf{GPU Version ($f_{p}^{h}$, $\gamma$, $\Delta$)}} \\ \hline \hline
& \textbf{arxiv} & \textbf{reddit} & \textbf{products} & \textbf{papers} & \textbf{arxiv} & \textbf{reddit} & \textbf{products} & \textbf{papers} \\ \hline
2 & 0.35, 0.9995, 128 & 0.50, 0.995, 64 & 0.25, 0.995, 128 & 0.50, 0.95, 256 & 0.25, 0.995, 32 & 0.35, 0.995, 32 & 0.50, 0.995, 32 & 0.50, 0.95, 512 \\ \hline
4 & 0.35, 0.95, 32 & 0.50, 0.95, 32 & 0.50, 0.995, 64 & 0.35, 0.995, 256 & 0.50, 0.995, 32 & 0.50, 0.995, 256 & 0.50, 0.995, 32 & 0.25, 0.995, 512 \\ \hline
8 & 0.15, 0.95, 256 & 0.35, 0.95, 64 & 0.35, 0.95, 64 & 0.50, 0.995, 1024 & 0.35, 0.995, 128 & 0.50, 0.95, 32 & 0.50, 0.9995, 16 & 0.50, 0.9995, 512 \\ \hline
16 & - & 0.25, 0.995, 16 & 0.25, 0.95, 32 & 0.50, 0.995, 512 & - & 0.50, 0.95, 16 & 0.35, 0.95, 32 & 0.50, 0.995, 512 \\ \hline
32 & - & - & - & 0.05, 0.95, 1024 & - & - & - & 0.15, 0.9995, 256 \\ \hline
64 & - & - & - & 0.15, 0.995, 1024 & - & - & - & 0.15, 0.9995, 256 \\ \hline
\end{tabular}
\label{tab:cpu_gpu_params}
}
\end{table*}

\subsubsection{Suitability for other models (GAT)}\label{sssec:eval-baseline-gat} 
\begin{figure}[!ht]
    \begin{subfigure}{0.8\columnwidth}
        \centering
        \includegraphics[width=\linewidth]{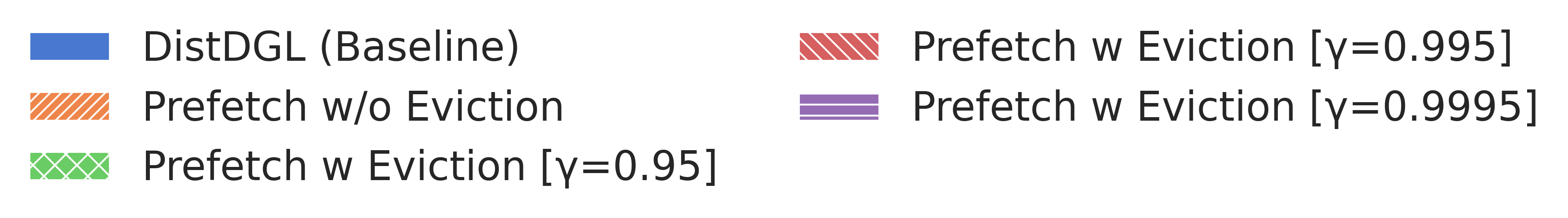}
        \label{fig:legend}
    \end{subfigure}
    \begin{subfigure}{0.24\textwidth}
        \includegraphics[width=\linewidth]{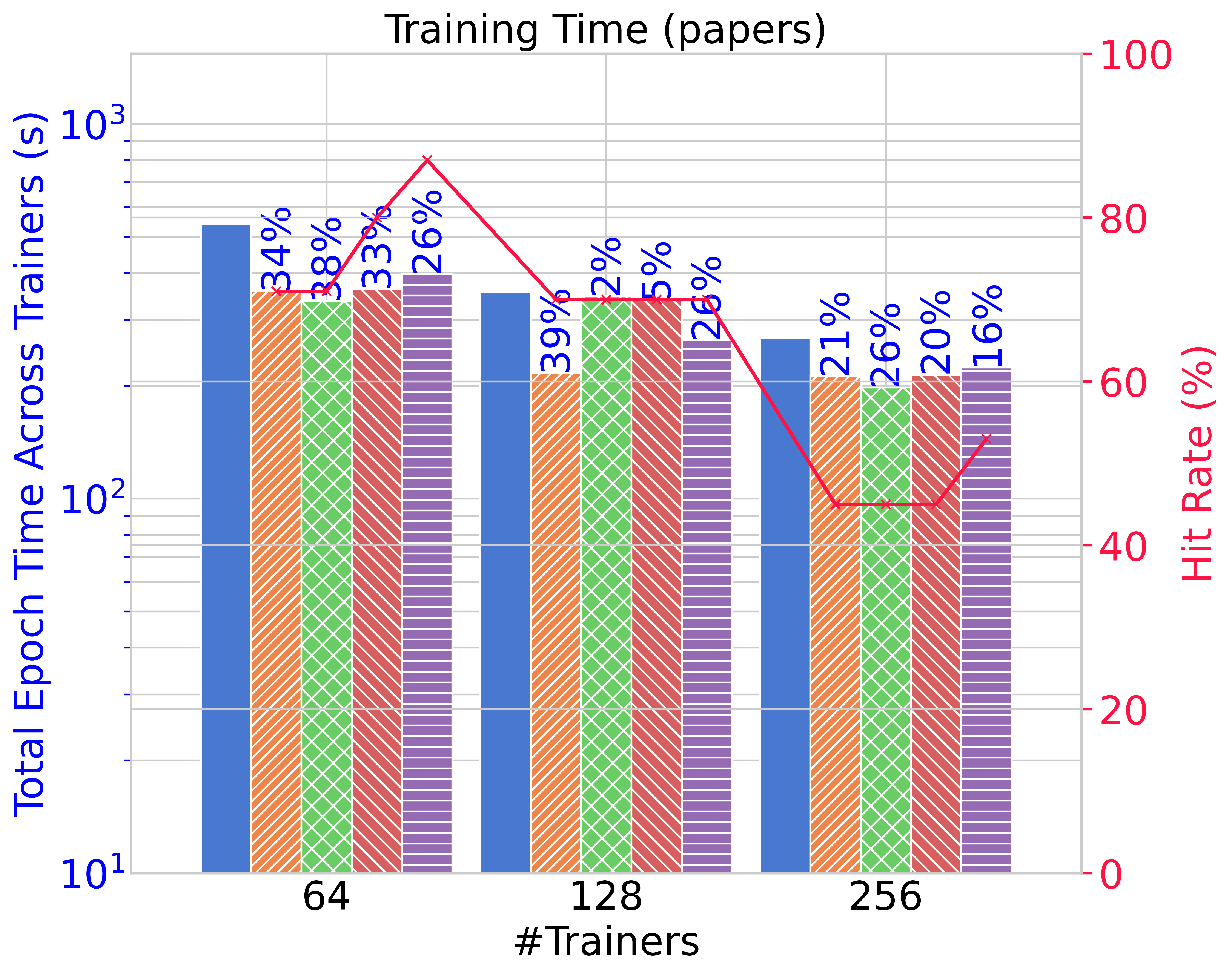}
        \caption{Papers-CPU}
        \label{fig:papers-cpu-gat}
    \end{subfigure}
    \begin{subfigure}{0.24\textwidth}
        \includegraphics[width=\linewidth]{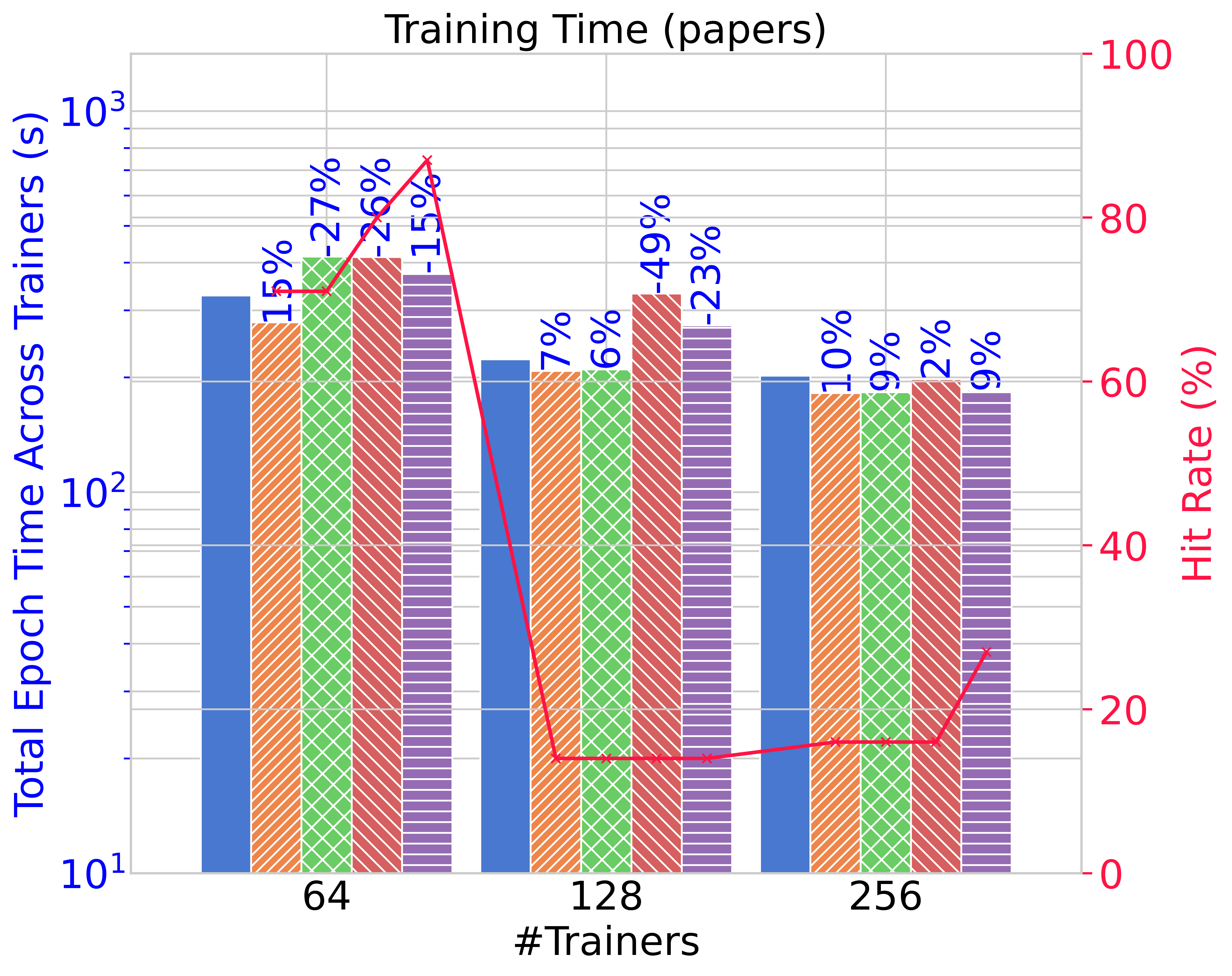}
        \caption{Papers-GPU}
        \label{fig:papers-gpu-gat}
    \end{subfigure}
    \caption{Training performance on GAT against DistDGL in papers across 2 versions (CPU, GPU) and up to 64 node configurations (4 trainers/node). [Y1: Lower is better; Y2: Higher is better]}
    \label{fig:baseline-gat}
\end{figure}

We further extend our evaluation to the Graph Attention Network (GAT) to determine if our prefetching scheme is adaptable to other GNN architectures. For this test, we configured the GAT model with two attention heads (the most that would fit in memory with the batch size of $2000$) with \texttt{NeighborSampler} as a sampler. We choose the largest input (papers) to train on $64-256$ trainers. As depicted in Fig. \ref{fig:baseline-gat}, our \emph{prefetch without eviction} approach provides improvements of up to 39\%  on CPU and 15\% on GPU. The eviction strategy further provides $5-8$ percent points on CPU when optimally configured. However, the GPU version fails to provide any impact due to our inability to accommodate a more complex GAT with more than two heads within the GPU's memory constraints, which adversely affected overlap efficiency (more discussions in \S\ref{sssec:analysis-metric}). The flattening of hitrate is due to $50\%$ and $70\%$ decrease in number of minibatches processed per trainer (discussed in \S\ref{sssec:eval-baseline-hitrate} and shown in Table \ref{tab:num_halo}). Note that we always prioritize time over hit rate when choosing optimal $f_{h}^{p}, \Delta$ and $\gamma$ in each version (CPU/GPU), which leads to different final combinations of parameters in each version shown in Fig. \ref{fig:baseline-gat}. The effectiveness of our prefetching scheme is influenced more by the choice of sampler than GNN architecture. While the complexity of GNN architecture can affect overlap efficiency, the performance primarily hinges on how the sampler interacts with the \emph{Prefetcher}, thus being versatile across GNN architectures.

\subsection{Observations and Analysis}\label{ssec:analysis}
\subsubsection{Initialization Costs}\label{sssec:analysis-prefetcher-init}
\begin{figure}[ht]
    \centering
    \includegraphics[width=\columnwidth]{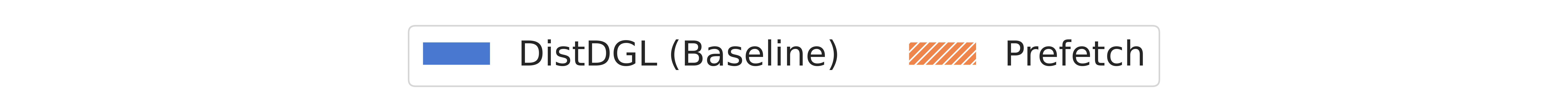}
    \includegraphics[width=\columnwidth]{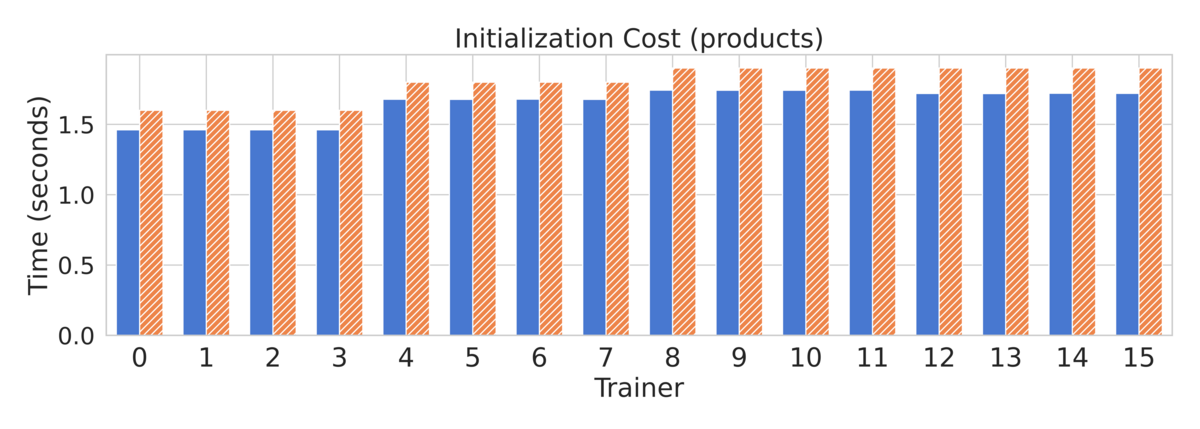}
    \includegraphics[width=\columnwidth]{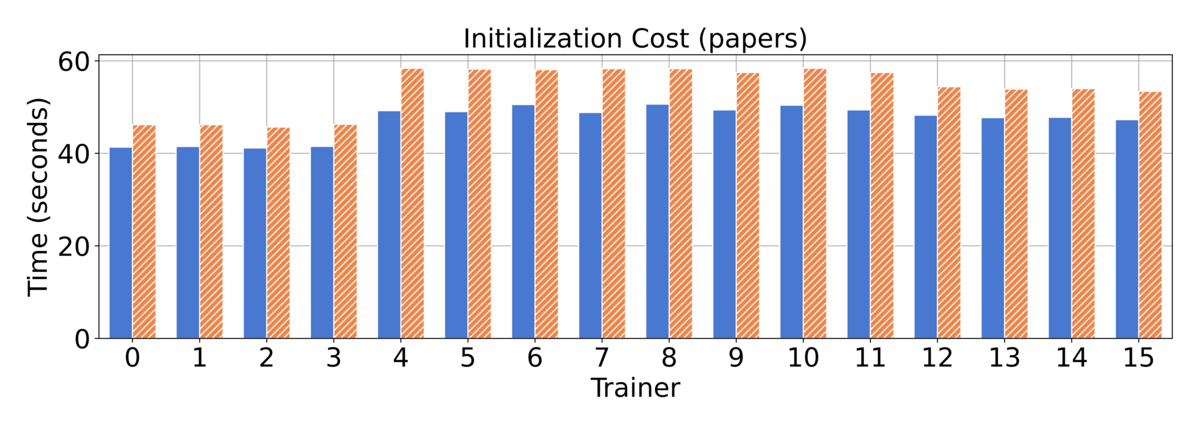}
    \caption{Initialization cost of products (top) and papers (bottom) on 4 CPU nodes. [Lower is better]}
    \label{fig:initialization-cost}
\end{figure}

The initialization of the prefetcher involves a one-time cost associated with selecting remote nodes, fetching their features, and populating the buffer with these features, as well as initializing the scoreboards $S_A$ and $S_E$ prior to training. The cost associated with this initialization phase for products and papers on 4 nodes is shown in Fig.~\ref{fig:initialization-cost} and is less than 1\% of the overall training execution time ($9\%-15\%$ more than DistDGL).

\begin{figure*}[t]
    \centering
    \begin{subfigure}{\textwidth}
        \centering
        \includegraphics[width=0.8\linewidth]{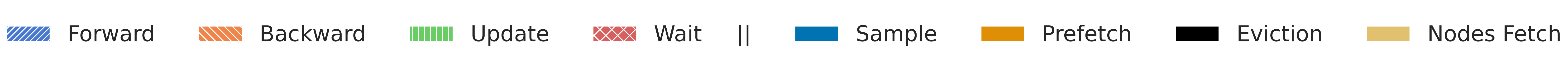}
        \label{fig:legend}
    \end{subfigure}
    \begin{subfigure}{0.245\textwidth}
        \includegraphics[width=\linewidth]{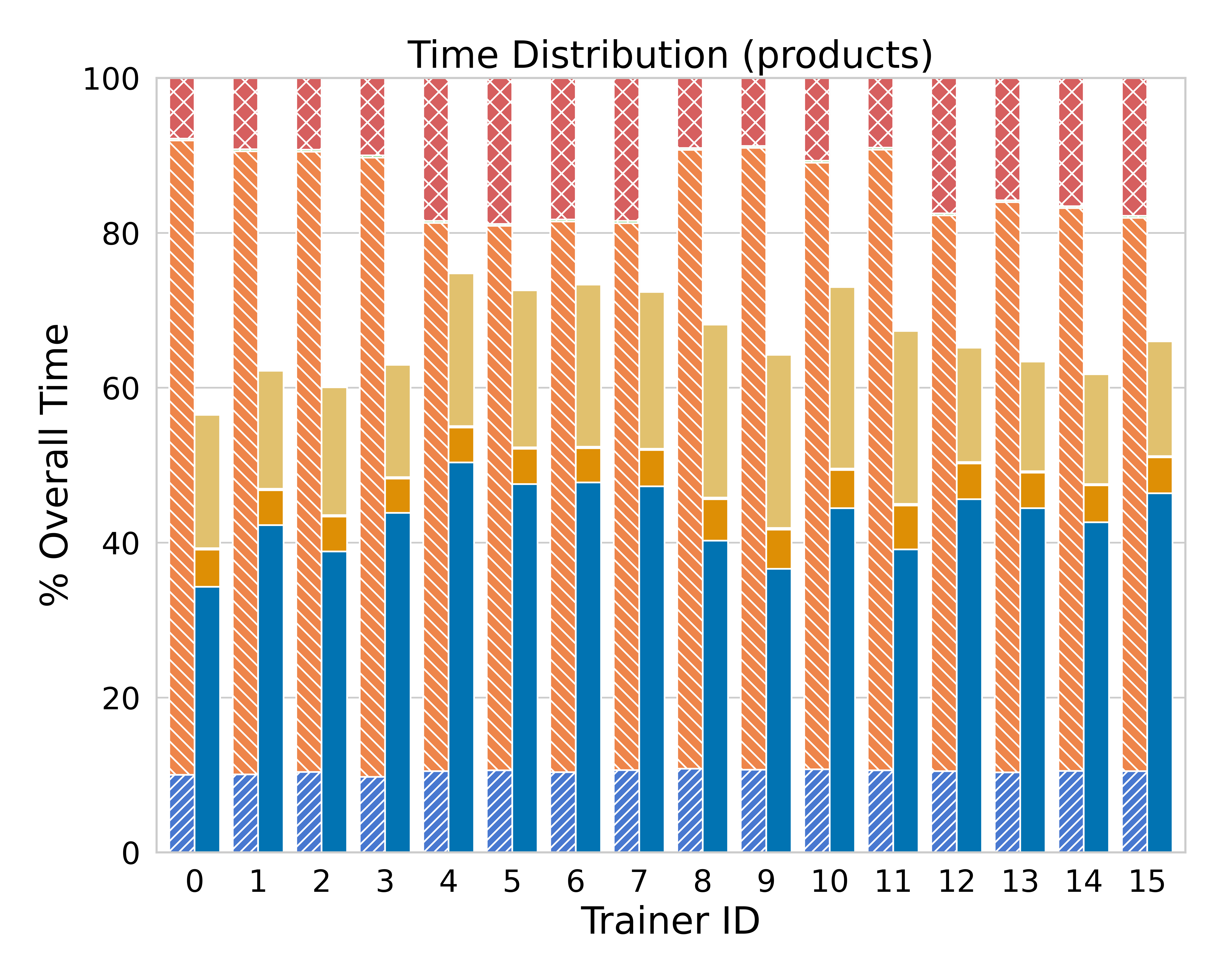}
        \caption{Products-CPU}
        \label{fig:prod-break-cpu}
    \end{subfigure}
    \begin{subfigure}{0.245\textwidth}
        \includegraphics[width=\linewidth]{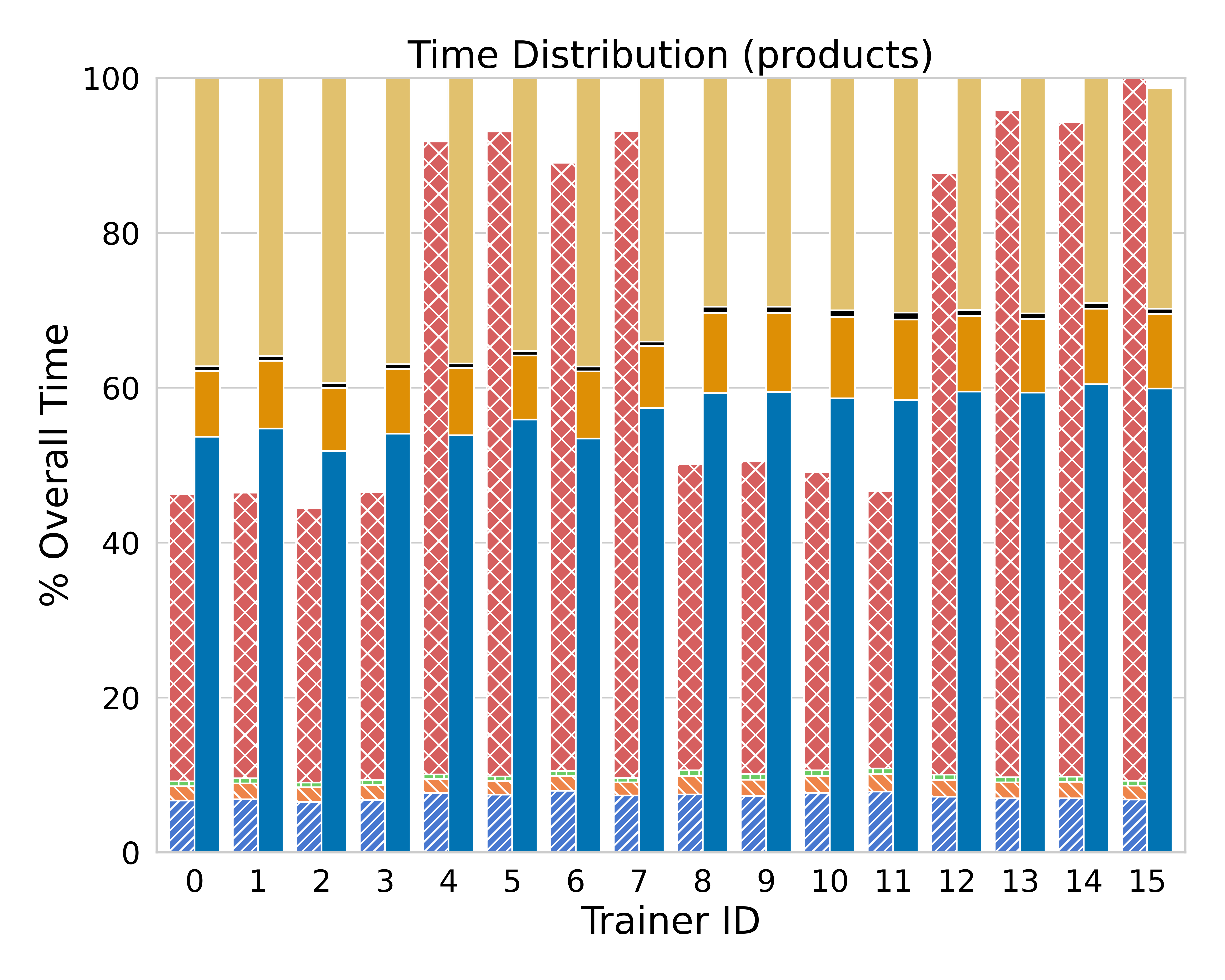}
        \caption{Products-GPU}
        \label{fig:pap-break-cpu}
    \end{subfigure}
    \begin{subfigure}{0.245\textwidth}
        \includegraphics[width=\linewidth]{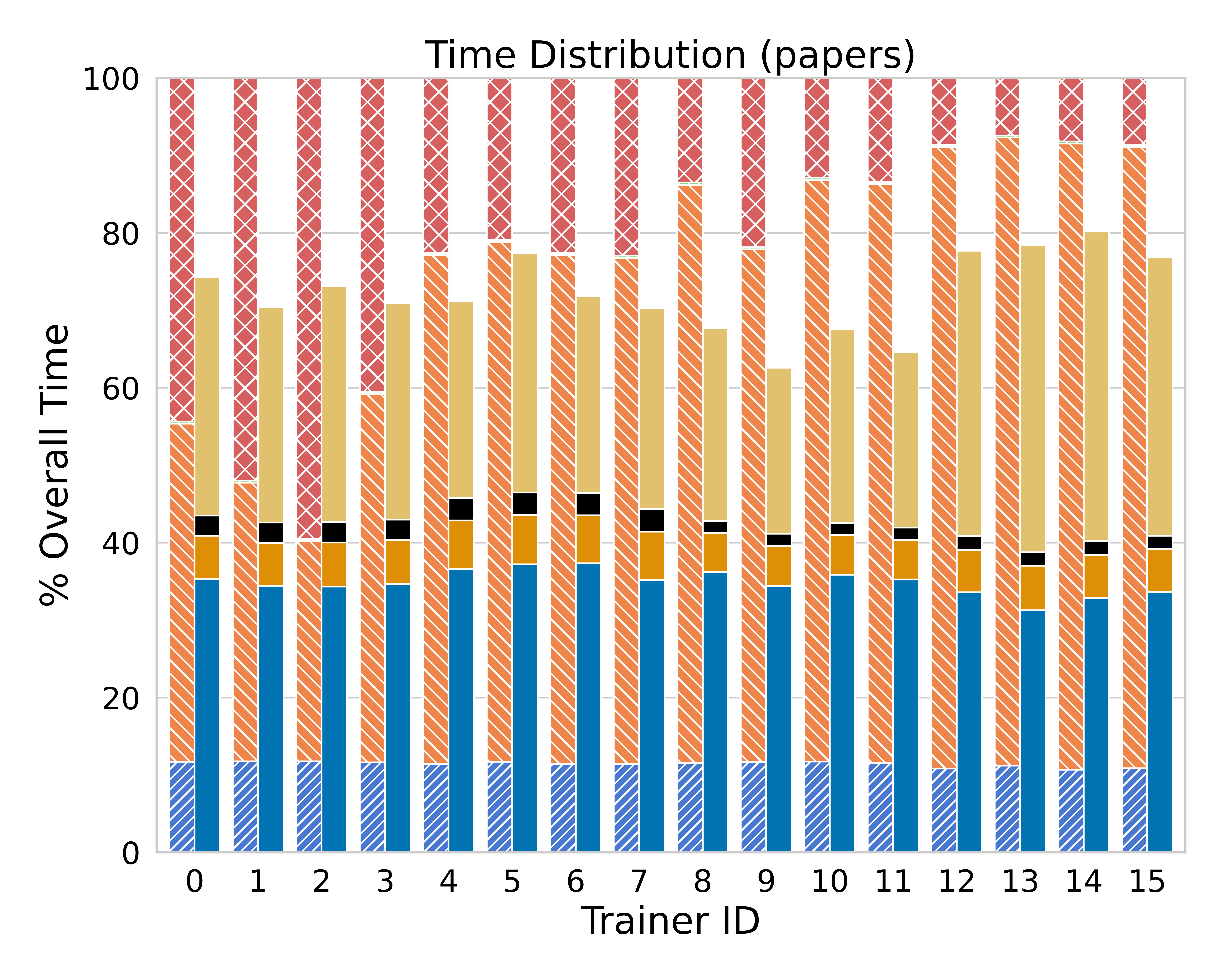}
        \caption{Papers-CPU}
        \label{fig:prod-break-gpu}
    \end{subfigure}
    \begin{subfigure}{0.245\textwidth}
        \includegraphics[width=\linewidth]{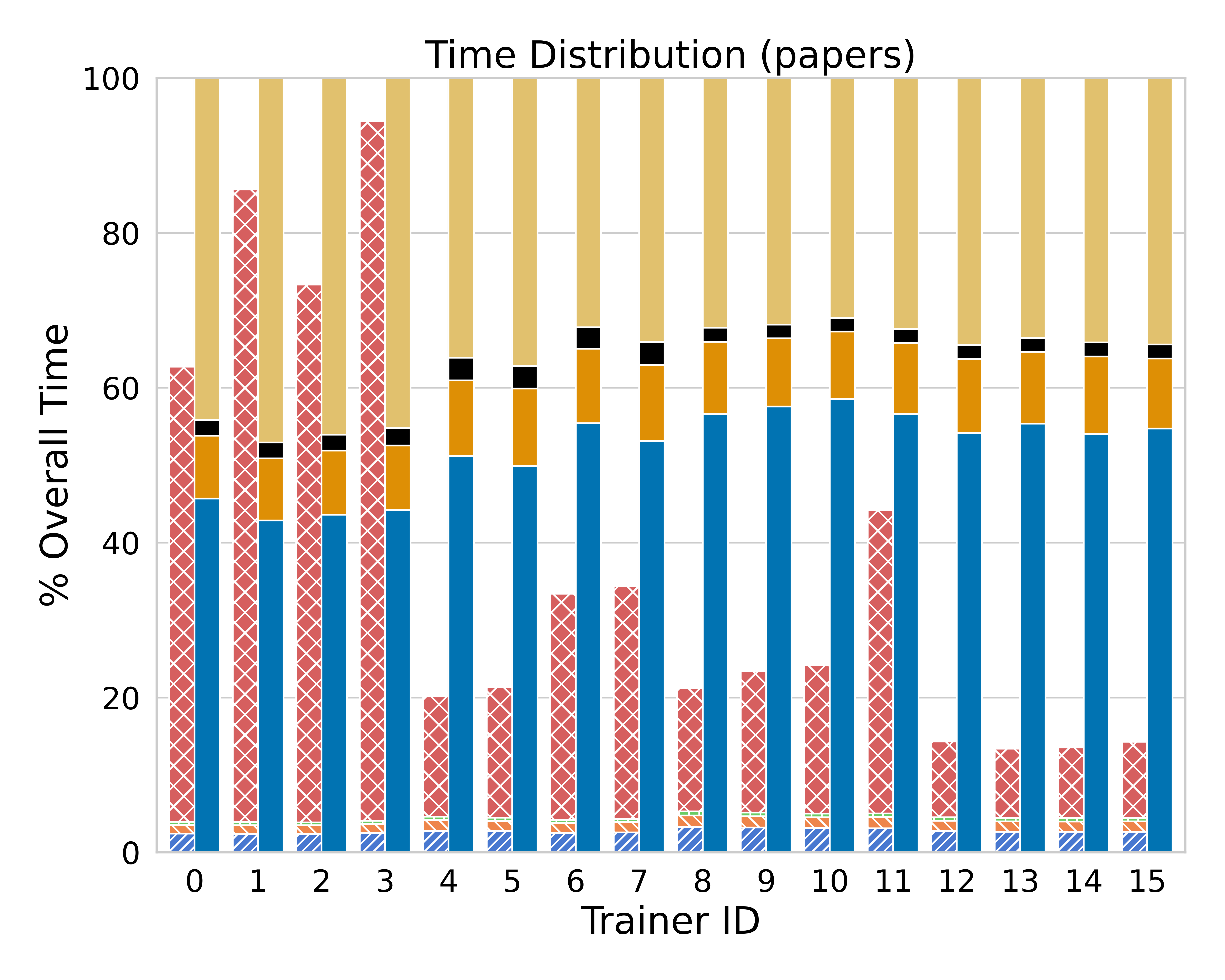}
        \caption{Papers-GPU}
        \label{fig:pap-break-gpu}
    \end{subfigure}
    \caption{Component-wise training time breakdown for products and papers on 4 CPU\slash GPU nodes using prefetching. Consecutive bars correspond to the overlap of \emph{current minibatch training} and \emph{next minibatch preparation}. In the baseline DistDGL version, the consecutive bars are stacked. Note that in (a), the eviction percentage is not clearly visible because it is $<1\%$ of the overall execution time.}
    \label{fig:breakdown}
\end{figure*}

\subsubsection{Overlap Efficiency}\label{sssec:analysis-overlap}
Overlap efficiency is the percentage of training time spent waiting for the preparation of the next minibatch to complete. This provides insights into the overlap between computation and data preparation, highlighting potential inefficiencies where the trainer PE is stalled, awaiting necessary data. Fig.~\ref{fig:breakdown} shows the overlap efficiency for CPU and GPU versions using papers and products inputs on 4 nodes. CPU versions consistently achieve 100\% overlap efficiency due to relatively longer training duration, while the GPU versions achieve $60\%-70\%$ overlap efficiency. This explains the greater improvements seen on CPUs compared to GPUs (especially for low \#nodes); with increasing \#nodes or trainers, the relative performance gains from prefetching between CPU\slash GPU versions are similar, as shown in Fig.~\ref{fig:baseline}.

\subsubsection{Eviction Analysis}\label{sssec:analysis-eviction}
\begin{figure}[h]
    \centering
    \includegraphics[width=0.5\columnwidth]{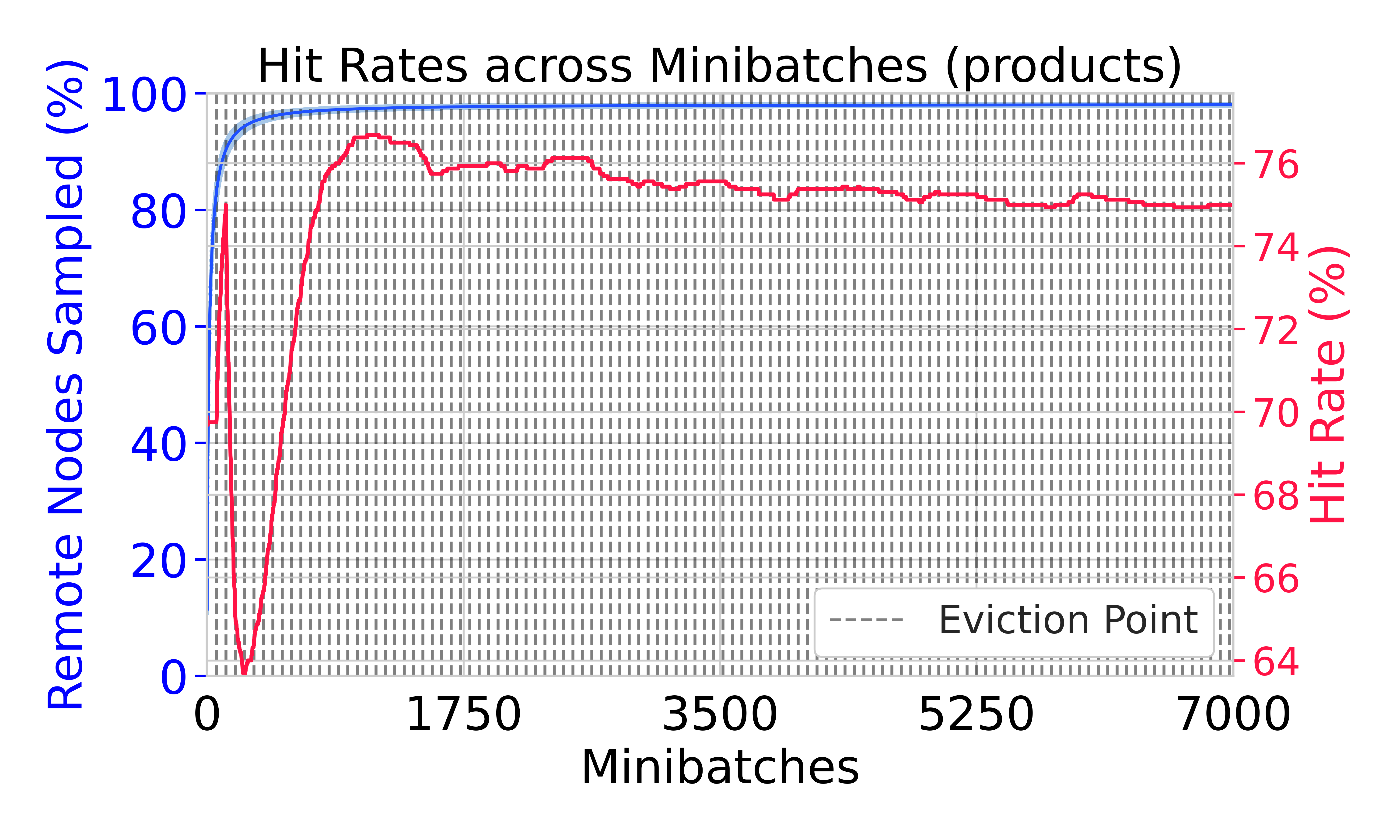}%
    \includegraphics[width=0.5\columnwidth]{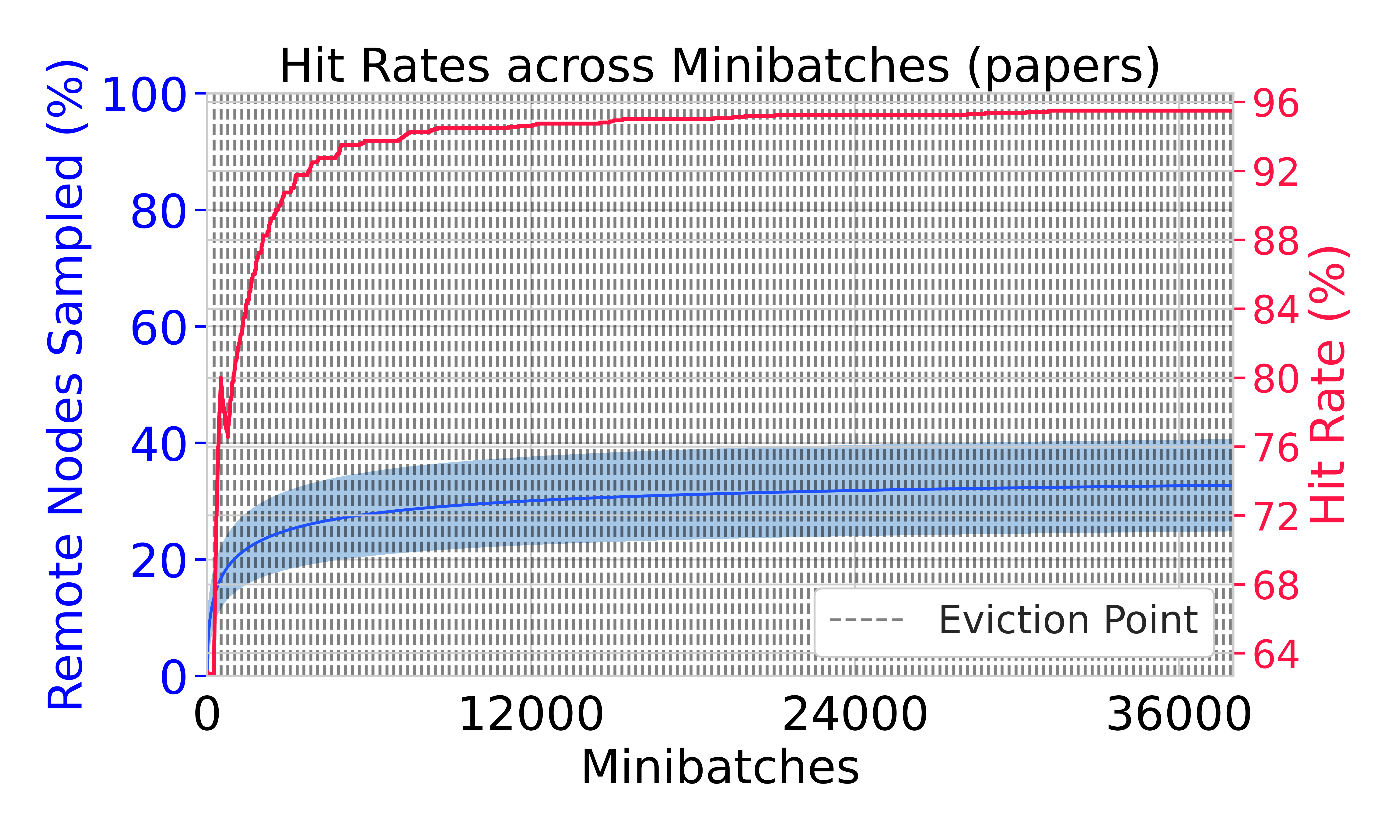}
    \caption{Progression of hit rate and percent of total remote nodes sampled across minibatches for products (left) and papers (right) on 4 CPU nodes. The blue shaded area represents variance among sampled nodes per trainer, while vertical dashed lines indicate eviction points as minibatches increase.}
    \label{fig:hitrate}
\end{figure}
The \emph{Prefetcher} populates the buffer primarily using two mechanisms: (1) a graph topology-based (degree-based) method for initializing and (2) a score-based approach for maintenance during training. Fig.~\ref{fig:hitrate} provides a comprehensive view of the eviction scheme with two contrasting cases. For the smaller input (i.e., products), the sampler returns more remote nodes (about 90\%) than local (due to relatively smaller-sized partitions), whereas, for the larger papers input (about 26$\times$ more edges than products), the partition sizes are relatively larger, leading to comparatively less amount of remote nodes being sampled (about 40\% of the total halo nodes in the current partition). We train for a longer duration (1000 epochs) to observe the hit rate trajectory. As the training progresses, the hit rate continues to rise at each eviction point ($\Delta$), ultimately plateauing at around 95\% in papers and 75\% in products. This suggests that the eviction scheme works in the long run by maintaining (through eviction and replacement) frequently sampled remote nodes, as indicated by the stability of (high) hit rates across minibatches. 

\subsubsection{Hit Rate and Performance}\label{sssec:analysis-metric}
The hit rate, which tracks the frequency of finding sampled remote nodes in the buffer, usually correlates with improved training performance (as we have seen in most instances of \S\ref{sssec:eval-baseline-gpu} and \S\ref{sssec:eval-baseline-cpu}). However, a high hit rate does not always yield significant performance gains. Our performance improvements are solely driven by two factors: 1) achieving overlap between training of the current minibatch and preparation of the next minibatch (``perfect overlap'' is the best case), and 2) retaining a nontrivial amount of remote nodes in the prefetch buffer which are subsequently sampled during future minibatches. If we achieve a perfect overlap and subpar hit rate, we will still break even in terms of performance (see discussion in \S\ref{ssec:method-model}), also empirically observed (i.e., declining hit rate, but 15\% performance improvements) for products and reddit datasets in Fig.~\ref{fig:baseline}. However, if the situation is reversed, i.e., we observe a high hit rate but see no overlap (when the $\frac{t_{RPC}}{t_{DDP}}$ in Equation~\ref{eqn:prebase} is less than 1), implying relatively higher training overhead than communication, then hit rate may not correlate with performance. We believe this scenario to affect some GPU instances of GAT, as shown in Fig.~\ref{fig:baseline-gat}, which exhibits a 10\% spike in hit rate and performance degradation relative to the baseline by a similar amount. 

\subsubsection{RPC Communication Analysis}\label{sssec:analysis-rpc}
\begin{figure}[!ht]
    \centering
    \includegraphics[width=0.9\columnwidth]{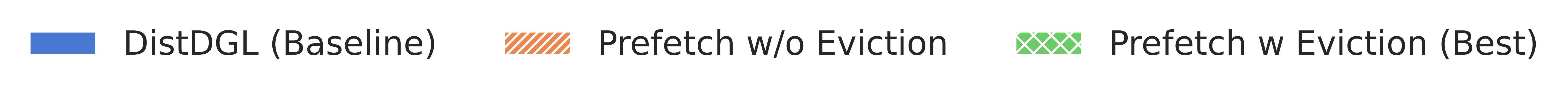}
    \includegraphics[width=0.9\columnwidth]{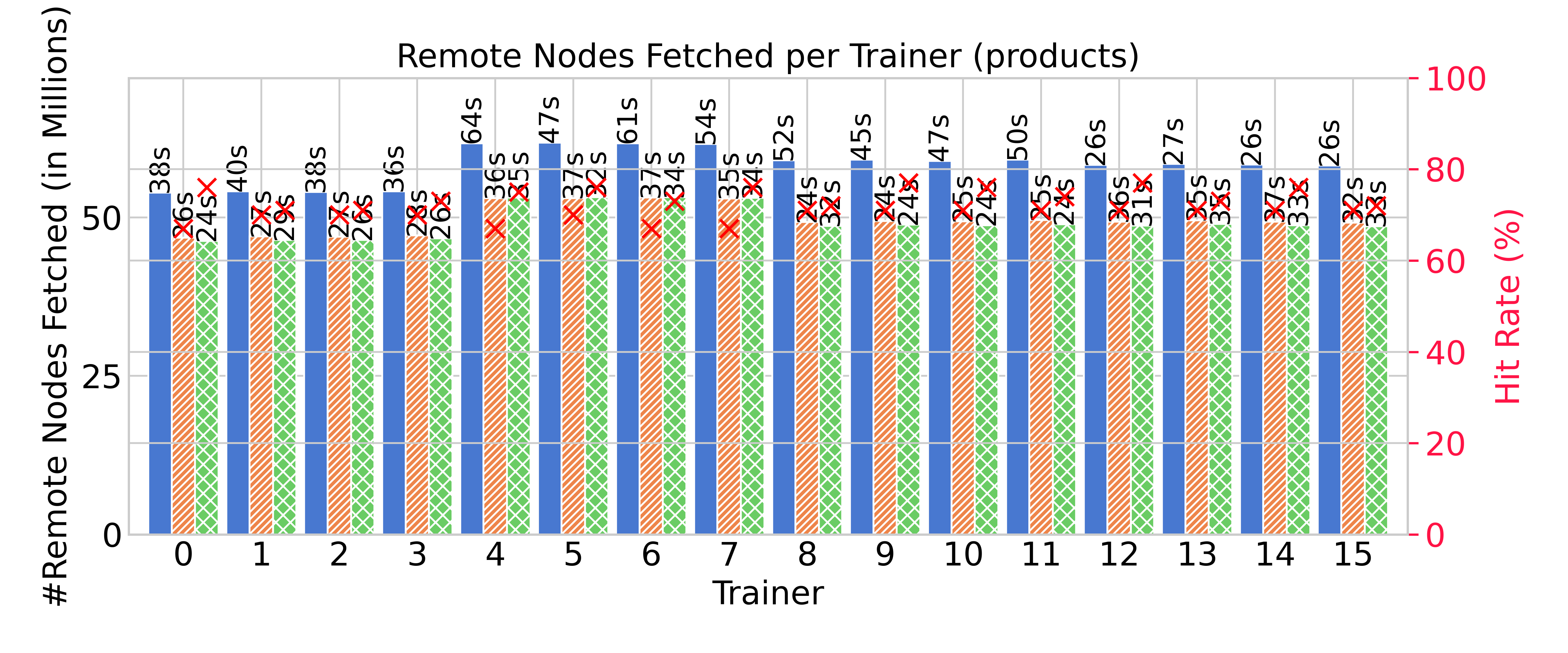}
    \includegraphics[width=0.9\columnwidth]{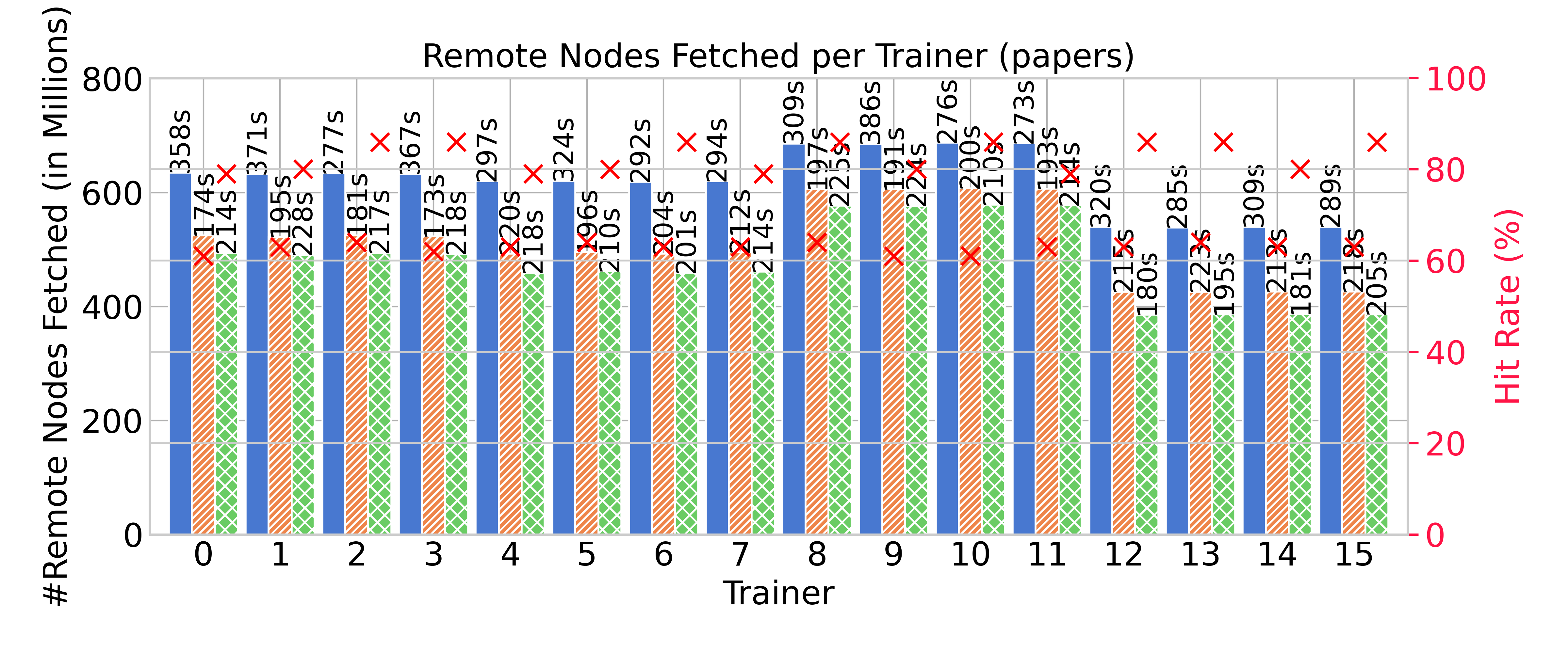}
    \caption{The number of remote nodes fetched with prefetching compared to baseline DistDGL during the training of products (top) and papers100M (bottom) on 4 CPU nodes using optimal $f_{h}^{p}, \Delta$ and $\gamma$. Annotations represent the communication time. [Lower is better in both cases]}
    \label{fig:rpc}
\end{figure}

We evaluate the impact of our prefetching approach on reducing the number of remote nodes requested per trainer. Minimizing this is crucial for reducing communication over time and improving overall efficiency. Fig. \ref{fig:rpc} shows a reduction of $23\%$ in papers and $15\%$ in products across 16 trainers with optimal $f_{p}^{h}, \Delta, \gamma$ and proves that our method effectively decreases remote node requests, even after accounting for the additional remote nodes requested during the replacement of evicted nodes. Further, we define the communication time to be the duration from the moment all remote node features are requested by a trainer from remote KVStores to the time they are received by the trainer. Since DistDGL overlaps communication time with local node lookup (Equation \ref{eqn:baseline} Part 2), the time stalled for communication: 
\begin{equation}
    t_{\text{communication}} = t_{RPC} - t_{Copy} = t_{(\text{send\_request} - \text{receive\_features})} - t_{Copy}
\end{equation}
Communication takes up to $29\%$ of the training time in products and up to $32\%$ in papers in 4 nodes, which our method reduces by around $44\%$ and $50\%$, respectively.

\subsubsection{Analysis of $f_{p}^{h}$, $\gamma$ and $\Delta$}{\label{sssec:analysis-gamma}}
\begin{figure}[!ht]
    \centering
    \begin{subfigure}{\columnwidth}
        \includegraphics[width=\columnwidth]{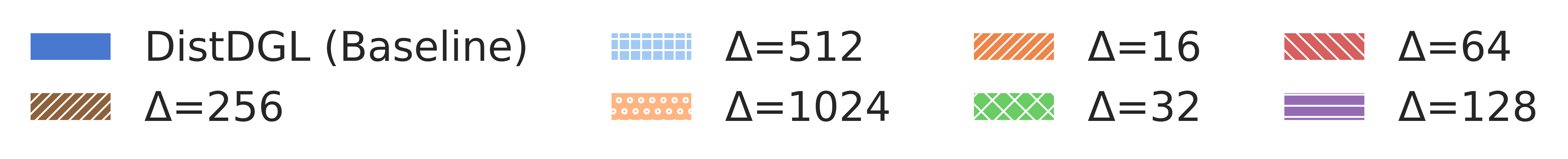}
        \label{fig:subplot1}
    \end{subfigure}
    \begin{subfigure}{0.49\columnwidth}
        \includegraphics[width=\columnwidth]{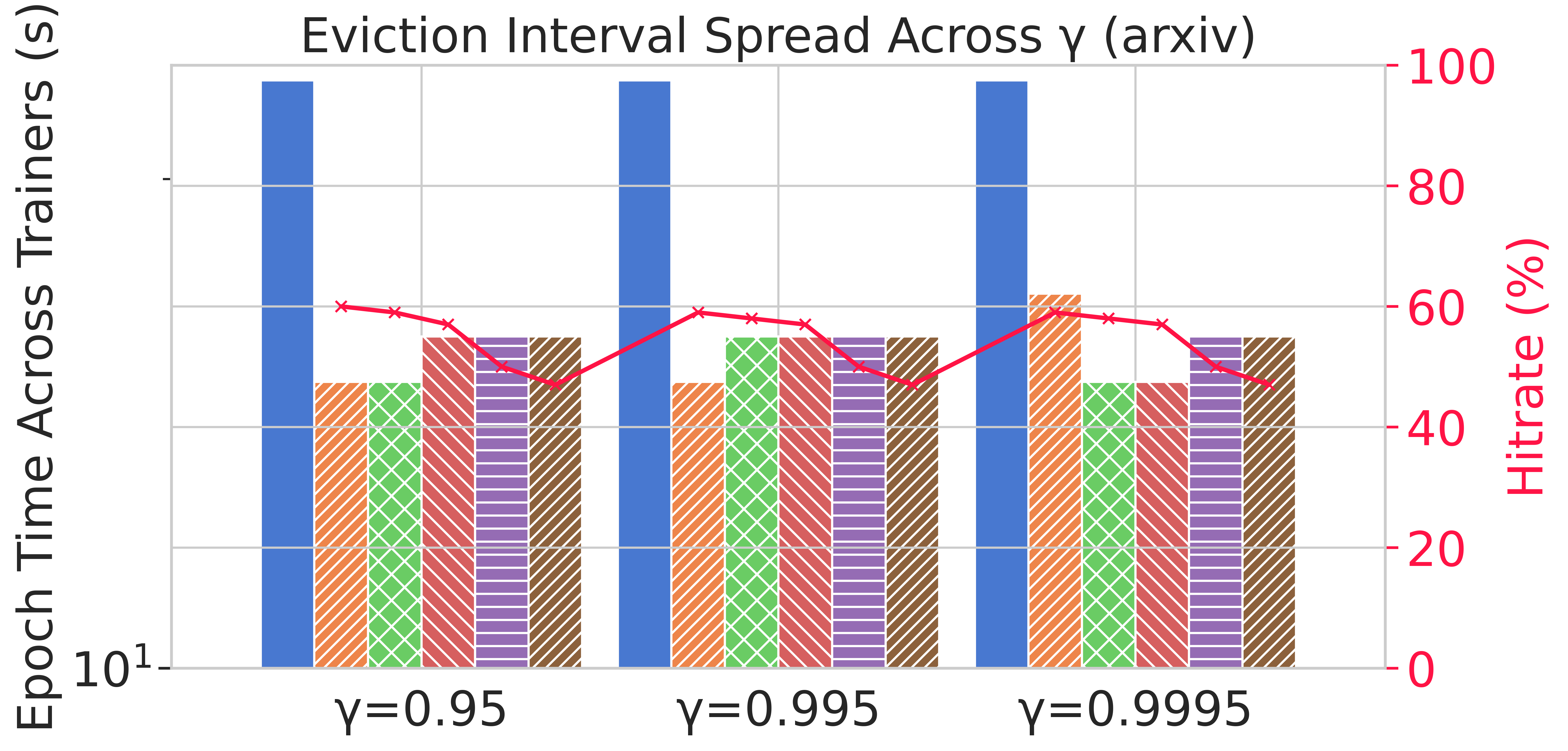}
        \label{fig:subplot1}
    \end{subfigure}
    \begin{subfigure}{0.49\columnwidth}
        \includegraphics[width=\columnwidth]{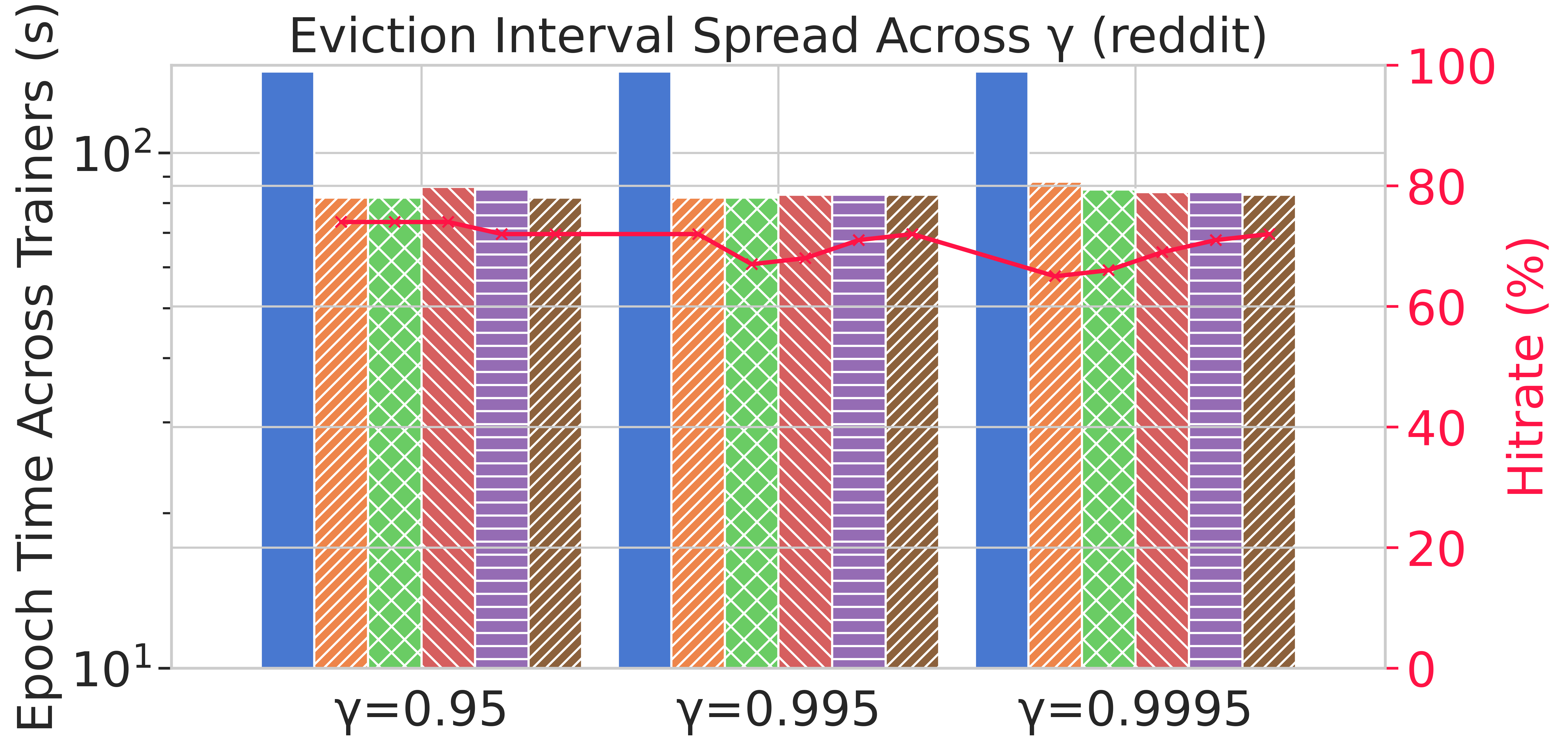}
        \label{fig:subplot1}
    \end{subfigure}
    \begin{subfigure}{0.49\columnwidth}
        \includegraphics[width=\columnwidth]{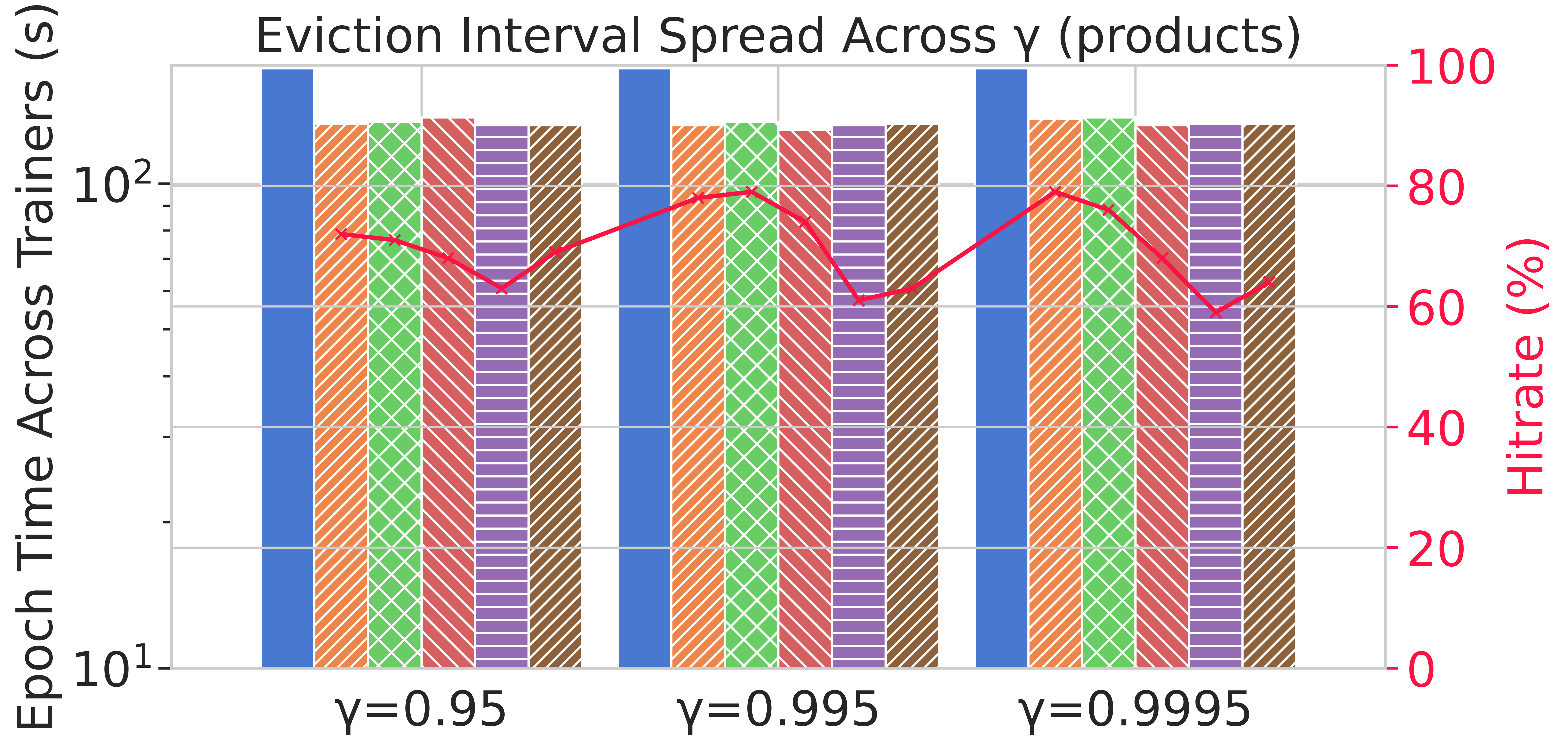}
        \label{fig:subplot1}
    \end{subfigure}
    \begin{subfigure}{0.49\columnwidth}
        \includegraphics[width=\columnwidth]{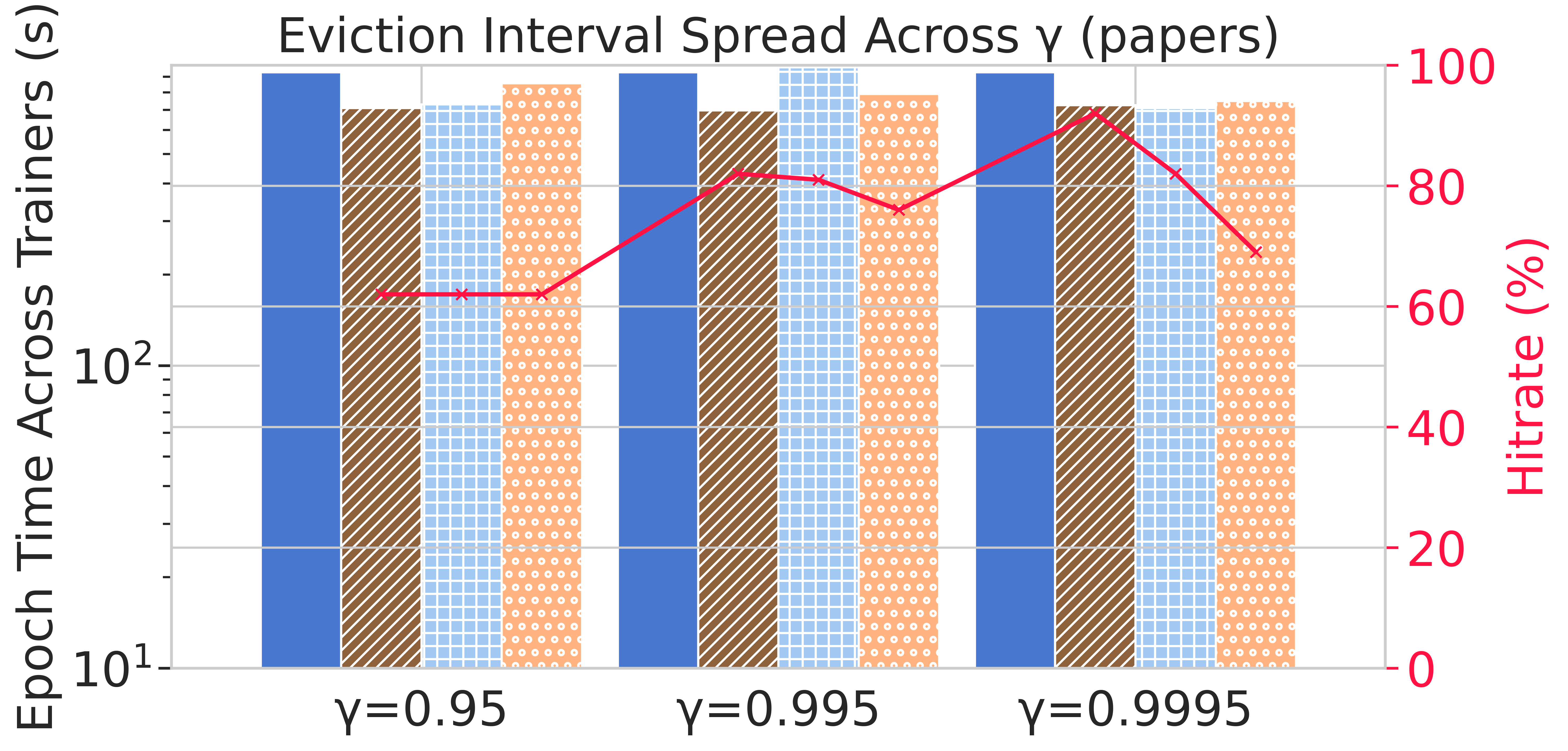}
        \label{fig:subplot1}
    \end{subfigure}
    \caption{Varying eviction interval ($\Delta$) per decay factor ($\gamma$)}
    \label{fig:period-spread}
\end{figure}

\begin{figure}[!ht]
    \centering
        \includegraphics[width=0.5\columnwidth]{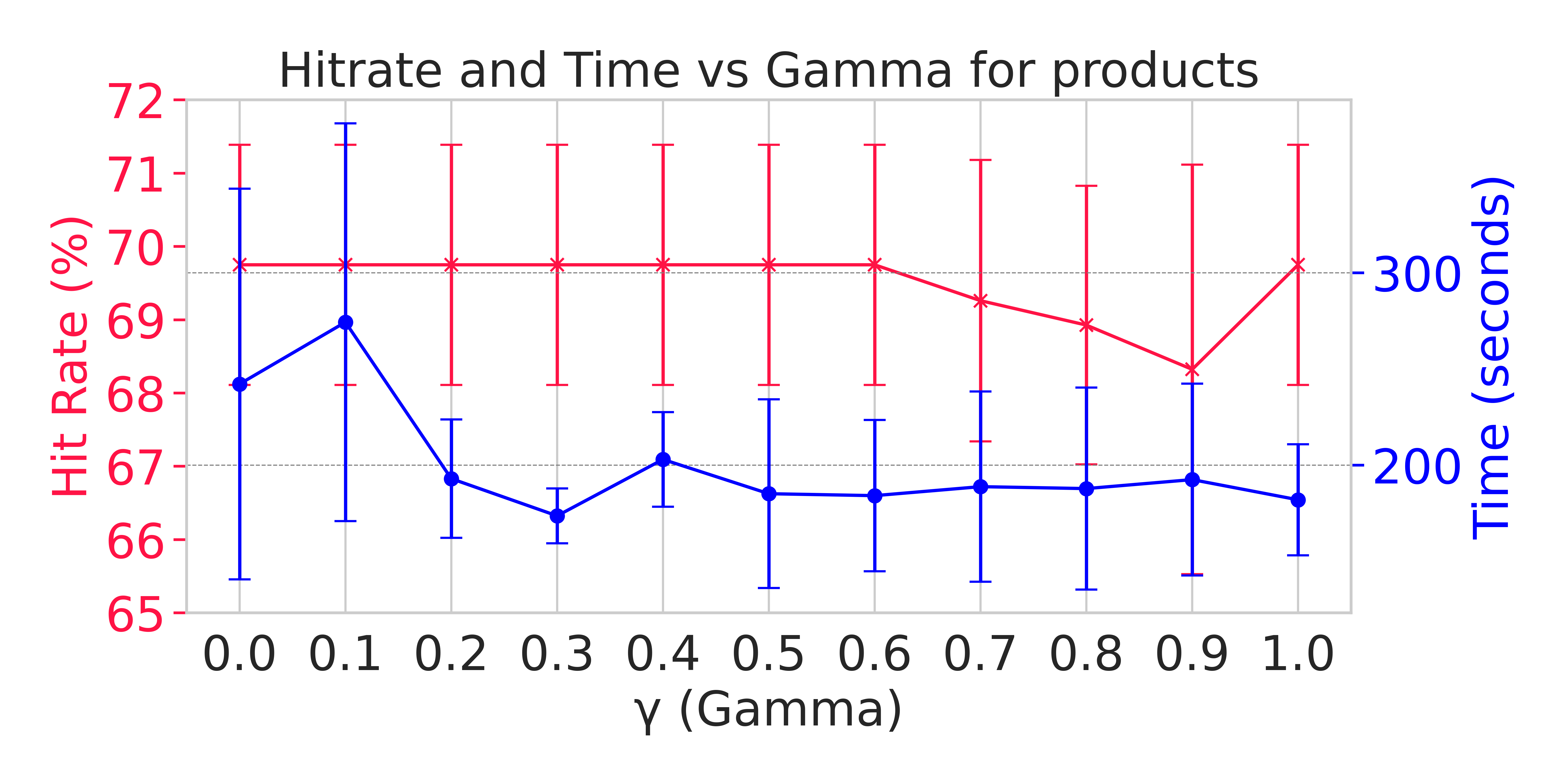}%
        \includegraphics[width=0.5\columnwidth]{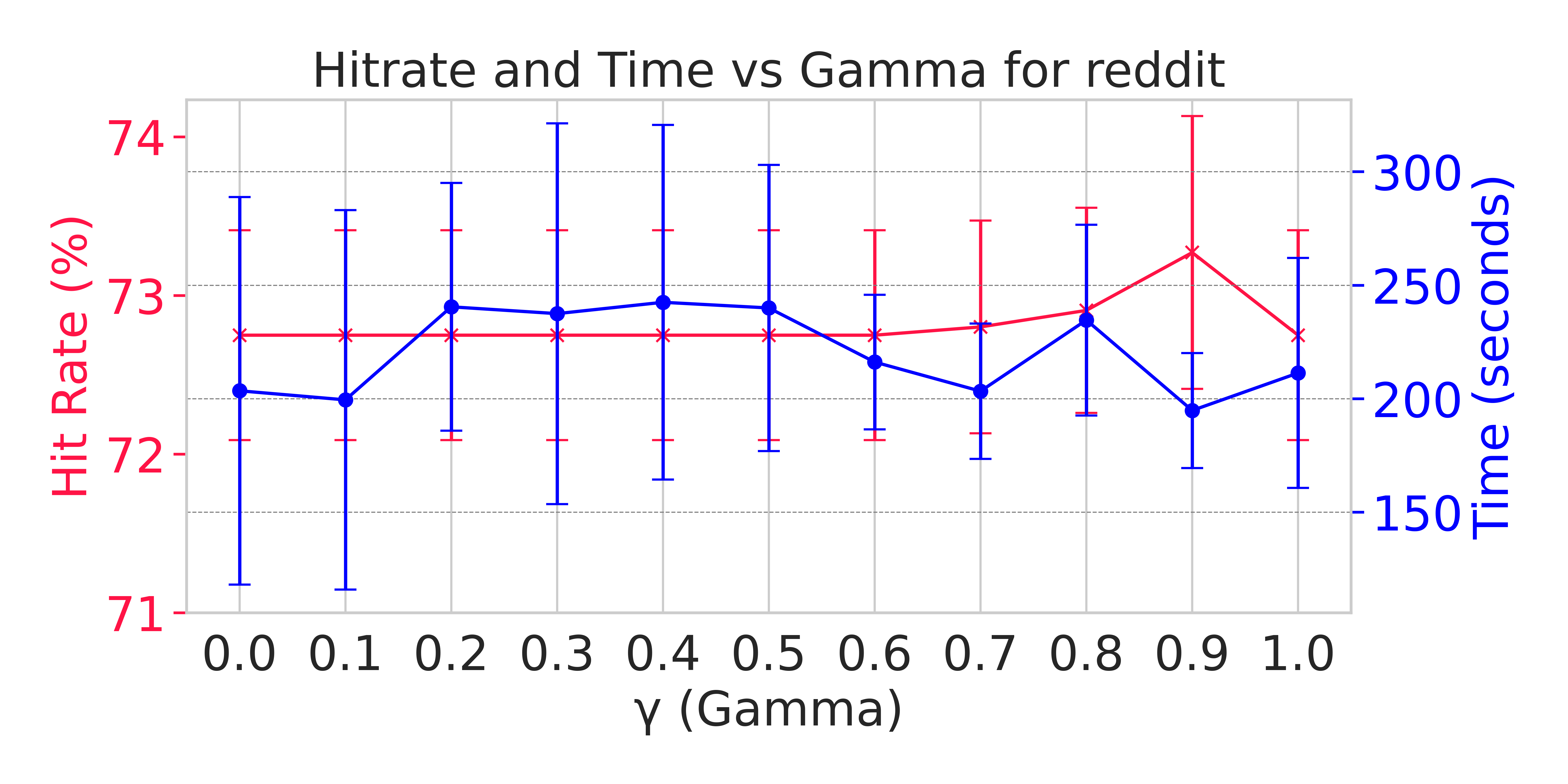}
        \caption{Varying decay factor ($\gamma$) across different intervals ($\Delta$)}
        \label{fig:gamma-analysis}
\end{figure}

As $f_{p}^{h}$ directly influences the buffer size, increasing it leads to higher memory usage (although peak memory consumption is bound by training, which is unchanged) and lookup overhead. However, a larger buffer also allows for more remote nodes to be maintained on a minibatch, which reduces the need for frequent evictions thereby reducing the communication cost, trading off memory with performance. We found different $f_{p}^{h}$ values to be optimal for various inputs on CPU/GPU platforms (Table \ref{tab:cpu_gpu_params}). The choice of the interval ($\Delta$) also affects prefetching-related computation overheads. When a higher interval is used, the eviction score ($S_E$, Algorithm~\ref{algo:prefetch-evict}) of relatively large number of (unused) nodes in the buffer drops below the threshold ($\alpha$), prompting to fetch replacements in bulk (instead of evicting fewer nodes per threshold period, which ultimately affects the eviction overhead, as discussed in \S\ref{ssec:method-tradeoff}). Sometimes, the sampler starts to re-sample nodes that were previously evicted. 
Fig.~\ref{fig:period-spread} demonstrates the effects of different eviction intervals ($\Delta$) for our choices of $\gamma$s on the execution time and the hit rate on 4 nodes. We also investigate the performance and hit rate trajectories for a range of $\gamma$ (0--1) on Fig.~\ref{fig:gamma-analysis} across the trainers on 4 (CPU) nodes. The error bars depict the ranges of the intervals ($\Delta$) considered between 16--1024. We observe that $\gamma\geq0.9$ (low decay) has a bigger hit rate spread across intervals (also corroborated in Fig.~\ref{fig:tradeoff}) yielding the best hit rates while also retaining relatively better execution time performance, which supports our choices of $\gamma$ for the baseline experiments (see \S\ref{ssec:eval-baseline}).
\subsubsection{Memory Usage}\label{sssec:analysis-memory}
\begin{figure}[!ht]
\centering
    \includegraphics[width=0.8\linewidth]{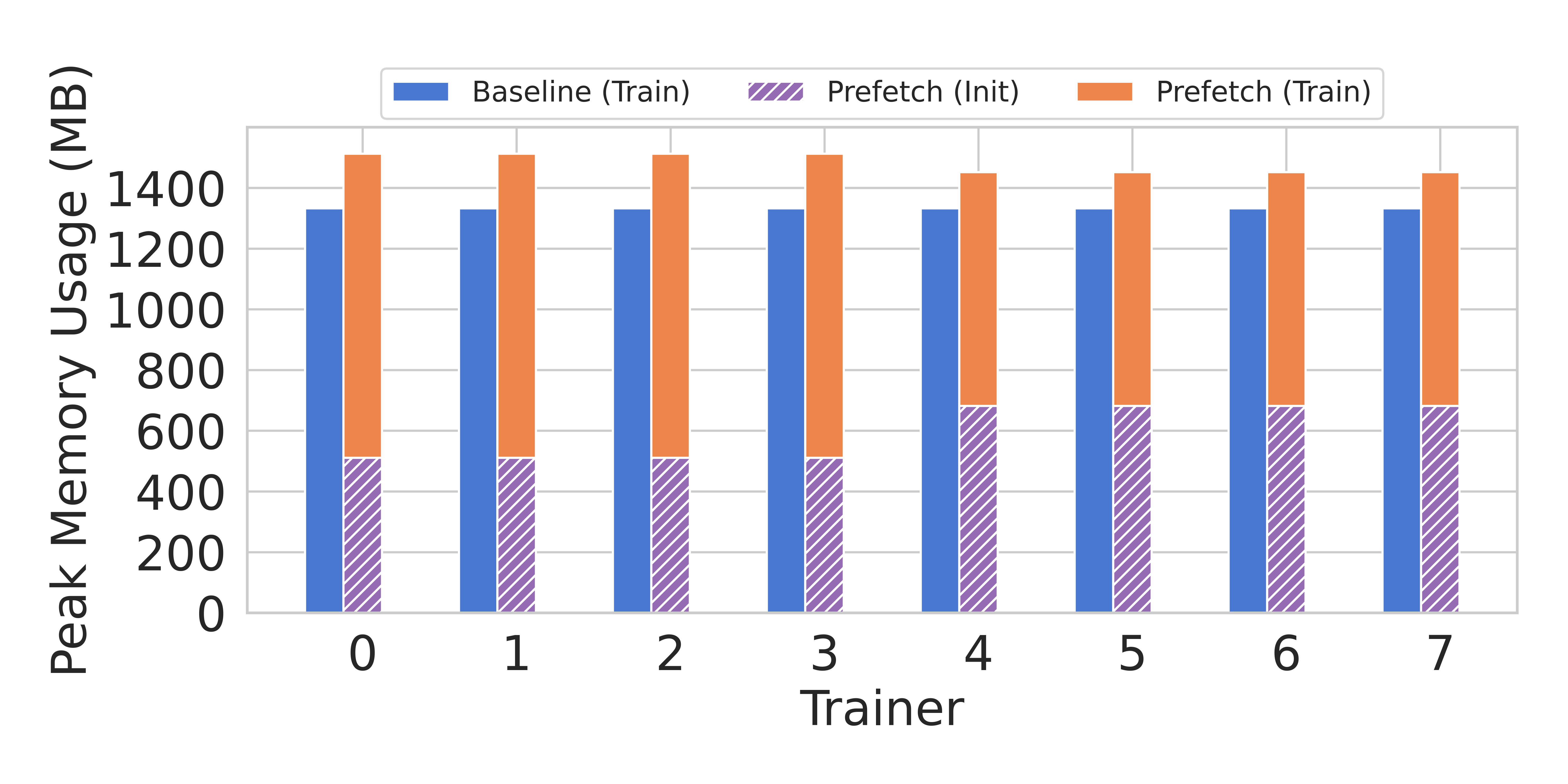}
    \caption{Peak memory usage of papers on 2 CPU nodes with $f_{p}^{h}=0.5$, $\Delta=1$ and $\gamma=0.95$ (extreme case) over 2 epochs.}
    \label{fig:mem}
\end{figure}

We use \emph{tracemalloc}\footnote{\url{https://docs.python.org/3/library/tracemalloc.html}} to collect memory allocations of baseline DistDGL and DistDGL updated with our prefetching scheme. We deliberately consider a memory-intensive configuration by evicting nodes at every minibatch from the prefetch buffer, which contains $50\%$ of the highest-degree remote nodes in each partition ($f_{p}^{h}=0.5$). Fig. \ref{fig:mem} demonstrates peak memory usage during initialization and training. 
Baseline DistDGL exhibits minimal allocations during initialization (about $0.11$MB, thus excluded from the figure), whereas the prefetch version takes about 500MB\slash trainer during initialization, mainly comprising of buffer and scoreboards. During training, there is a mild difference in the peak memory usage of baseline DistDGL vs. prefetch, about $10\%$ extra, due to the frequent eviction process (we see no evidence of a higher peak memory for the prefetching scenarios in Fig.~\ref{fig:baseline}). 

\section{Summary}\label{sec:summary}
Minibatch training on partitioned huge graphs leads to adversarial communication patterns, which undermines the efficiency of training. The proposed prefetch and eviction scheme can preempt communication overheads by maintaining relevant remote nodes' features locally, asynchronously preparing the next minibatch with concurrent training. Further, we implement our scheme within the state-of-the-art Amazon DistDGL framework with the intent of community adoption. Our extensive evaluations on 4 diverse OGB datasets and two architectures (GraphSAGE and GAT) highlight about $15--40\%$ improvements in the end-to-end training performances on contemporary CPU/GPU systems. 

Ongoing advances in compute node configurations (i.e., higher number of cores, memory capacity, and bandwidth) will allow us to consider larger prefetch buffers and engage more processing cores for preparing future minibatches, further optimizing the data movement over the network. Although ``perfect overlap'' is the best-case scenario and depends on several parameters, options to prefetch future minibatches can pave the way towards a sustainable ``perfect overlap'' model for various GPU-based configurations of GNN training.

\section*{Acknowledgments}
This research is supported by the U.S.\@ Department of Energy (DOE) through
  the Office of Advanced Scientific Computing Research's ``Orchestration for Distributed \& Data-Intensive Scientific Exploration'' and
  the ``Cloud, HPC, and Edge for Science and Security'' LDRD at Pacific Northwest National Laboratory.
PNNL 
is operated by Battelle for the DOE under Contract DE-AC05-76RL01830. This research is also supported by the National Science Foundation (NSF) under Award Number 2243775.

\bibliographystyle{plain}
\bibliography{references}

\end{document}